\documentclass[twocolumn,aps,prl,10pt,amsmath,amssymb,nofootinbib,showpacs,superscriptaddress,floatfix]{revtex4-1}
\pdfoutput=1

\DeclareFontFamily{U}{rcjhbltx}{}
\DeclareFontShape{U}{rcjhbltx}{m}{n}{<->rcjhbltx}{}
\DeclareSymbolFont{hebrewletters}{U}{rcjhbltx}{m}{n}

\usepackage{amsthm}
\newtheorem{definition}{Definition}
\usepackage{graphicx}
\usepackage{color}
\usepackage[usenames,dvipsnames]{xcolor}
\usepackage[colorlinks=true,linkcolor=Red,citecolor=Green,linktoc=page]{hyperref}
\usepackage{multirow}
\usepackage{float}
\usepackage{flushend}
\usepackage{balance}
\usepackage[varg]{txfonts}
\usepackage{ulem}
\usepackage{fancyhdr}
\usepackage{wasysym}

\begin{document}
\title{Networks as the fundamental constituents of the universe}

\author{C.\,A.\,Trugenberger}

\affiliation{SwissScientific Technologies SA, rue du Rhone 59, CH-1204 Geneva, Switzerland}
\affiliation{Division of Science, New York University Abu Dhabi, Abu Dhabi, United Arab Emirates}
\affiliation{Corresponding author: ca.trugenberger@bluewin.ch}

\begin{abstract}
Relations, represented by graph edges, are essential for modelling complex systems across various fields, including physics, computer science, and biology. They provide a universal framework for capturing interactions, dependencies, and pathways, with networks encoding these intricate connections. In this context, we review an approach that uses binary relations as the fundamental constituents of the universe, utilizing them as building blocks for both space and matter. This model, known as combinatorial quantum gravity, is defined by an ultraviolet continuous fixed point of a statistical model on random networks, governed by the combinatorial Ollivier-Ricci curvature, which acts as a network analogue of the Einstein-Hilbert action. The model exhibits two distinct phases separated by this fixed point, a geometric and a random phase, representing space and matter, respectively. At weak coupling and on large scales, the network organizes into a holographic surface whose collective state encodes both an emergent 3D space and the matter distributed in it. The Einstein equations emerge as constitutive relations expressing matter in terms of fundamental network degrees of freedom while dynamics in a comoving frame is governed by relativistic quantum mechanics. Quantum mechanics, however is an effective theory breaking down at the scale of the radius of curvature of the holographic network. On smaller scales, not only relativistic invariance is lost but also the Lorentzian signature of space-time. Finally, the manifold nature of space-time breaks down on the Planck length, where the random character of the fundamental network on the smallest scales becomes apparent. The network model seems to naturally encode several of the large-distance features of cosmology, albeit still at a qualitative level.
The holographic property of black holes arises intrinsically from the expander nature of random regular graphs. There is a natural mechanism to resolve the cosmological constant problem  and dark matter appears naturally as a metastable allotrope in the network fabric of space-time. In this model, both gravity and quantum mechanics are macroscopic statistical effects reflecting the free energy minimization of fundamental binary degrees of freedom. 
\end{abstract}
\maketitle

\section{Introduction}
This special issue is dedicated to discrete curvature measures and their applications. When discussing curvature in physics, the first concept that comes to mind is general relativity. Einstein's groundbreaking insight was to propose that space-time behaves like an elastic medium, capable of bending and stretching to mediate gravitational interactions. According to this framework, matter curves space-time, and this curvature, in turn, influences the motion of matter. However, the relativity of time inherent in this picture conflicts with quantum mechanics, which relies on a rigid, Newtonian notion of time \cite{kiefer, kuchar}. Furthermore, unless the relativity of time is abandoned at very short scales, general relativity is non-renormalizable \cite{horava}.

Most modern approaches to this problem assume that quantum mechanics is the more fundamental theory and seek to "quantize" gravity \cite{polchinski, as1, as2, loopqg, cdt1, cdt2, oriti, tensor1, tensor2, sasakura, causalsets, gorard, verlinde1, verlinde2, bianconient}. An alternative perspective, however, is to try to derive both general relativity and quantum mechanics from an even deeper underlying model based on binary degrees of freedom \cite{wheeler}. In this view, it is important to recognize that geometry is not limited to continuous manifolds; rather, it is a fundamental property of metric spaces in general. Graphs, for instance, possess their own intrinsic geometry through graph distance. Could it be that the fundamental building blocks of the universe are abstract relations—the edges of a graph? Note that this is very different from the standard computational trick of discretizing a manifold \cite{cdt1, cdt2}: here one admits metric configurations that have nothing to do with manifolds. We would also like to stress that (undirected) graphs can always be interpreted as quantum entanglement patterns \cite{zanardi}, although this is not the route we will pursue here. 

In recent years, significant progress has been made in the study of discrete geometry, particularly in defining purely combinatorial notions of curvature for general metric spaces and networks \cite{forman1, forman2, olli0, olli1, olli2, olli3, olli4}. This has inspired an approach to formulating general relativity on networks, using the combinatorial Ollivier-Ricci curvature as a discrete counterpart to the Einstein-Hilbert functional \cite{comb1, comb2, comb3}. The idea is to construct a statistical model on random networks (for a review see \cite{graphrev}), where the Boltzmann weight is determined by the Ollivier combinatorial curvature \cite{olli0, olli1, olli2, olli3, olli4}, and a dimensionless coupling controls the strength of fluctuations. This coupling serves as a precursor to Planck’s constant, with the fluctuations it generates giving rise to the familiar quantum behaviour—but only at large scales, within a specific phase of the model. Indeed, in this approach, Planck's constant cannot be a fundamental constant of nature since its definition requires the concepts of energy, a length scale and a speed, which are emergent. At the most fundamental level, the intrinsic fluctuations of binary degrees of freedom can only be governed by a dimensionless coupling. For related attempts to obtain manifolds from networks see \cite{bianconi2, bianconi3}. 

A key characteristic of the graph ensemble we consider is its ability to exist in two distinct phases. At weak coupling, the system is in a geometric phase, where the graph forms a tiling of a manifold. At strong coupling, however, it transitions into a random phase, where the defining properties of manifolds are lost. The transition from random to geometric phase is continuous and driven by the condensation of short cycles \cite{comb1, comb2, comb3}. In the random phase, the graph is locally tree-like, with few cycles and distances that scale logarithmically with the number $N$ of vertices, corresponding to an infinite Hausdorff dimension. In contrast, the geometric phase is rich in short cycles, with distances scaling as $N^{1/2}$ and the graph becomes the tessellation of a holographic surface. Due to the continuous nature of the transition, a fundamental scale—the Planck length—emerges naturally in the geometric phase. This scale, given by the correlation length, characterizes the typical size of clusters of the "wrong" phase that persist at finite coupling. These random phase clusters have Planck size and exhibit higher energy and curvature than the surrounding geometric phase, making them natural candidates for matter particles. In this model, both space and matter arise as different configurations of network edges (for a review see \cite{universe}). Moreover, higher-curvature (and higher-energy) allotropes of the space tiling provide an intuitive candidate for dark matter \cite{dark}. 

A key feature of quantum gravity, shared by nearly all proposed approaches, is that space-time has a two-dimensional (2D) structure at short scales \cite{horava, spectralcdt, reuter, verlinde1}. This can be understood as a manifestation of the holographic principle \cite{thooft, susskind} (for a review see \cite{bousso}): the fundamental degrees of freedom of gravity reside on a 2D holographic surface, while the higher-dimensional structure emerges at larger scales. Consequently, we focus on networks whose geometric phase corresponds to 2D surfaces. As we will demonstrate, the networks that minimize the free energy at finite coupling, when endowed with a uniform length scale $\ell$, form tilings of negatively curved surfaces.

To describe the dynamics of the model—that is, the fluctuations around the equilibrium configurations previously discussed—we must introduce a universal time parameter. As in the Ising model, there are essentially two types of fluctuations in the surviving clusters of “wrong” (random) phase at small but finite coupling. The first type alters the order parameter—here, the total number of short cycles defining the space tiling—and corresponds to the growth or shrinkage of the clusters \cite{binder}. The second type, akin to Kawasaki dynamics in the Ising model \cite{tersenghi, luck}, involves fluctuations at fixed order parameter and describes the diffusion of clusters within their surrounding space. A full analysis of these network fluctuations remains an open and challenging problem. However, the diffusion component can be effectively studied in the continuum limit, where clusters are modeled as point particles diffusing in a space of negative curvature.

Diffusion in Riemannian spaces with negative curvature exhibits remarkable properties \cite{prat, kendall, hsu, hsubrief, hsubook, arnaudon}. At large scales, it separates into two parts: a dominant ballistic (deterministic and inertial) component, and a subdominant random correction \cite{anker}. Remarkably, up to exponentially small errors, this ballistic component admits a dual description as time-like geodesic motion in a Lorentzian de Sitter space with positive curvature \cite{eff1, eff2}. In essence, inertial trajectories in Riemannian spaces of negative curvature become equivalent, at large scales, to time-like geodesics in Lorentzian spaces of positive curvature. Both exhibit divergence—one due to hyperbolic geometry, the other due to Hubble expansion. Importantly, ballistic diffusion has a maximum velocity, determined by the fastest speed at which information can propagate through the network \cite{luck}. This speed is essentially set by the relaxation time of a single edge.

Focusing on this dominant deterministic behaviour, we effectively obtain a (1+1)-dimensional Lorentzian holographic screen, where one dimension is identified with the universal Newtonian time governing diffusion and becomes, thus an emergent geometric time \cite{eff1, eff2}.  This is the same (1+1)-dimensional screen obtained in most quantum gravity models. However, in the present case, to fully capture the dynamics, we must also account for the random component of diffusion. When this is included, two significant changes occur. First, Newtonian time no longer coincides with geometric time. Secondly, and perhaps more surprisingly, an additional spatial dimension emerges. Comoving observers on large scales perceive the random component in a Euclidean 3D space with distances inherited from the hyperbolic distances on the 2D holographic screen \cite{anker, ledrappier, cllt, davies}. In the dual Lorentzian perspective, this random component is quantum mechanics in three spatial dimensions, with the upper velocity bound identified as the Lieb-Robinson limit \cite{eff1, universe}. This large-scale dynamics can be ``lifted" to relativistic quantum dynamics in a space-time which comoving observers perceive as (1+3)-dimensional de Sitter space. In the continuum limit, the equations defining matter in terms of fundamental network degrees of freedom take the form of the Einstein equations. In this model, these are not dynamical equations but, rather constitutive equations. The dynamics is encoded in the fundamental network degrees of freedom. 

The overall picture of this model is as follows. Quantum gravity is non-perturbatively defined by an ultraviolet fixed point for a holographic network governed by the combinatorial Ollivier Ricci curvature. At small coupling and large scales, the network state encodes both an emergent 3D space and a matter distribution on it. The Einstein equations emerge as constitutive equations defining matter in terms of microscopic fundamental degrees of freedom. The dynamics of matter in a comoving frame is relativistic quantum mechanics. However, quantum mechanics is an emergent theory which breaks down on the scale of the radius of curvature of the holographic network. On small scales, not only relativistic invariance is lost but also the Lorentzian signature of space-time. As pointed out originally by Polyakov \cite{polyakovhouches}, the very existence of antiparticles is a strong hint of an underlying Euclidean structure of space-time and this emerges naturally here. The manifold nature of space-time breaks down on the Planck length, where the random character of the fundamental network on even smaller scales appears. 

The network model appears to reproduce naturally a number of cosmological large-scale features, although so far only at a qualitative level. The holographic character of black holes follows directly from the expander nature of random regular graphs. There is a natural mechanism to avoid the cosmological constant problem and, finally, dark matter arises in a natural way as a metastable allotrope within the network fabric of spacetime.

In this model, gravity and quantum mechanics arise at large scales from the interplay of curvature energy and entropy of microscopic network degrees of freedom, with both matter and geometric manifolds emerging at the free energy minima. Contrary to \cite{verlinde1, verlinde2} and to a recently proposed network model \cite{bianconient} which shares some key ideas with \cite{comb1, comb2, universe}, the principle driving the emergence of geometry is not entropy maximization but, rather, free energy minimization, where energy is the network analogue of the Einstein-Hilbert action. To explain this difference, a brief comment on the statistical mechanics of networks is in order. In most network models addressed in the literature, there is no underlying microscopic principle to determine the most probable network realizations, apart a set of overall constraints, like, e.g., the number of edges or the degree sequence. As a consequence, the only principle for choosing a graph probability distribution is the information-theoretic maximization of entropy. This results in a Boltzmann distribution, with an effective energy given by the constraints, which, in the network literature, is called the exponential random graph model \cite{newman}. In the present case of ``general relativity on networks", however, the situation is reversed: the fundamental quantity is an energy functional which defines the Boltzmann distribution according to statistical mechanics principles, the network curvature. The most probable network configurations at a given coupling constant are then determined by minimizing the corresponding free energy, rather than maximizing entropy under constraints.

In this contribution we shall review this model of combinatorial quantum gravity, with a particular focus on the graph-theoretic aspects concerning discrete curvature notions and their applications.

\section{Combinatorial curvature}
The classical Einstein equations of general relativity represent the stationary points of the Einstein-Hilbert action. At the quantum level, this action is expected to govern the fluctuations of space-time manifolds. However, the corresponding sum over continuous geometries is plagued by the proliferation of ultraviolet divergences. Two common approaches to address this issue are: introducing new physics at short scales, as in string theory, or assuming a fundamental discrete structure at small scales, as is the case in most other models. An alternative approach is to embrace both. Indeed, geometry is not limited to continuous manifolds; every metric space, including discrete ones, inherently possesses a distance measure, allowing the development of discrete geometry. Additionally, a notion of curvature can be defined on such discrete spaces, enabling the formulation of general relativity on discrete metric spaces \cite{comb1, comb2, comb3}. Since every finite discrete metric space can be represented as a network \cite{metric}, we will focus on the concept of combinatorial curvature in network structures. As continuum Ricci curvature is associated with a direction on a manifold, its combinatorial analogue must be tied to an edge of the network. Several notions of combinatorial Ricci curvature have been advanced. Here we will focus on two of the most popular candidates and their relation. 

\subsection{Forman curvature}
A first discrete measure of graph curvature was introduced by Forman \cite{forman1, forman2} using the topological constructs of CW (Closure-finite, Weak) cell complexes.
In contrast to Regge calculus \cite{regge} and similar curvature measures used in the discretization of manifolds \cite{cdt1, cdt2}, this curvature is not defined geometrically but only combinatorially. Its formulation draws upon an analogy with identities developed  by Bochner \cite{bochner} concerning the decomposition of the Riemannian-Laplace operator on the space of $p$-forms, $\Omega^p(M)$ defined for a manifold $M$. This decomposition yields a covariant derivative and a curvature correction known as the Bochner-Weitzenb\"ock identity, whose discrete form is used to define the Forman-Ricci (FR) curvature.

Intuitively, a CW complex is a space made by gluing together simple building blocks of different dimensions (points, segments, disks, balls etc.) along their boundaries.
Formally, CW complexes (for a review see \cite{hatcher}) are constructed from $p$-cells ($p$ denoting the dimension of the cells). A $d$-dimensional CW complex is made by gluing $p \leq d$ complexes along shared faces. Thus, a $1$-complex, encompassing only nodes, i.e. $0$-cells and links, i.e. $1$-cells, is a graph. The original FR of an edge, thus, can depend at most on the degrees of the two vertices defining the edge. While this is a computationally simple construct, it is by far too coarse to capture the ``geometry" of networks. 

The Forman curvature can by ``augmented" by considering graphs as $2$-complexes, with graph cycles assumed to bound $2$-cells \cite{sreejith}. These proposals, however focus typically only on triangles as $2$-cells \cite{samal}, which is still a too strong limitation when considering a discretization of Ricci curvature. Ricci curvature involves second derivatives of the metric, which is itself a measure of infinitesimal distances. The graph analogue of a metric is an edge, while the discrete graph equivalent of derivates up to second order involves up to the second-nearest neighbours of the two vertices defining this edge. Therefore, a more appropriate ``enhancement" of FR of graphs, treated as $2$-complexes, should involve triangles, squares and pentagons \cite{tee}. This is essentially the discrete version of locality on graphs. 

The boundary of a $p$-cell is the ensemble of $(p-1)$-cells that ``bound" the cell. For a $1$-cell, a graph edge, the boundary is the collection of the two vertices defining the edge. A general $p-$cell, $\alpha_p$ is a proper face of a $(p+1)$-cell $\beta$ if it is a member of the boundary set of $\beta$, and we write $\alpha_p < \beta_{p+1}$. A $p$-cell CW complex $M$ over $\mathbb{R}^p$, is defined formally as a collection of cells $\alpha_q \ , q \in \{0,\dots,p\}$, such that any two cells are joined along a common proper face, and all faces are contained in the cell complex.

A further important concept about cell complexes is the definition of the neighbours of a given $p$-cell \cite{forman1, forman2},
\begin{definition}
\label{def:neighbor}
	Let $\alpha_1$ and $\alpha_2$ be $p$-cells of a complex $M$. $\alpha_1$,$\alpha_2$ are neighbours if:
	\begin{enumerate}
		\item $\alpha_1$ and $\alpha_2$ share a $(p+1)$-cell $\beta$ such that $\beta > \alpha_1$ and $\beta > \alpha_2$, or
		\item $\alpha_1$ and $\alpha_2$ share a $(p-1)$-cell $\gamma$ such that $\gamma < \alpha_1$ and $\gamma < \alpha_2$.
	\end{enumerate}
\end{definition}
\noindent We can thus partition the set of neighbours of a cell into parallel and non-parallel ones. Two $p$-cells $\alpha_1$,$\alpha_2$ are parallel neighbours if one but not both of the conditions in Definition \ref{def:neighbor} are true, in which case we write $\alpha_1 \parallel \alpha_2$. 

Using these concepts, one can define a series of maps $\mathcal{F}_p : \alpha_p \rightarrow \mathbb{R}$, for each value of $p$, 
\begin{equation}\label{eqn:fr-simple}
	\mathcal{FR}_p (\alpha_p) = \# \{ \beta_{(p+1)} > \alpha_p \} + \# \{ \gamma_{(p-1)} < \alpha_p \} - \# \{ \epsilon_q \parallel \alpha_p \} \mbox{,} 
\end{equation}
where $\epsilon_q$ is a $q$-cell that is a parallel neighbour of $\alpha_p$ and $q \neq p$. The symbol $\#$ is intended to denote the number of such cells satisfying the condition in braces. Essentially this definition computes $\mathcal{FR}_p (\alpha_p)$ as the number of $(p-1)$-cells that bound $\alpha_p$, plus the number of $(p+1)$-cells of which $\alpha_p$ is part of the boundary minus the number of parallel neighbours of $\alpha_p$. For $p=1$,  $\mathcal{FR}_p \left(e_{ij} \right) = \kappa^{FR}_{ij}$ defines the Forman curvature FR of edge $e_{ij}$. In this case, eq. \eqref{eqn:fr-simple} is particularly simple; the vertices and edges constitute the $0$- and $1$-cells and the closed loops in the graph constitute the $2$-cells.
In eq. \eqref{eqn:fr-simple} there is no arbitrary restriction to the length of cycles that are admissible as bounding a $2$-cell. The so-called ``augmented" FR is obtained by restricting these cycles to triangles. The geometrically more appropriate ``enhanced" FR $\kappa^{FR}$ \cite{tee} is obtained by truncating the expansion to cycles of maximum length $5$. 

Another very important property of $CW$ complexes is that of quasi-convexity \cite{forman2}, 
\begin{definition}
\label{def:qconvextiy}
	A $CW$-complex $M$ is quasi-convex if, for each pair of $( p +1)$-cells $\alpha_1$ and  $\alpha_2$ (viewed as topologically closed entities including their boundaries), 
	if $\alpha_1 \cap \alpha_2$ contains a $p$-cell $\beta$, then $\alpha_1 \cap \alpha_2 = \beta$. 
\end{definition}
\noindent In the case of graphs, viewed as $2$-complexes, this property reduces to the simple condition that any two cycles can share at most one edge (two edges share anyway always one link). Moreover, in view of the fact that only cycles of length up to 5 are relevant for local properties on the graph, we shall consider a reduced notion of quasi-convexity of graphs that imposes the condition of maximally one shared edge only for triangles, squares and pentagons. As we now show, this condition not only renders the computation of various combinatorial curvature measures tractable but it is a very physical condition representing a sort of ``incompressibility" of graphs; it was independently introduced in \cite{comb1} and \cite{comb2} where it was called ``hard core condition" or ``condition of independent short cycles". 

For quasi-convex graphs, satisfying the independent short cycles condition, the prohibitive expression eq. \eqref{eqn:fr-simple} reduces to the simple form 
\begin{equation}\label{eqn:exact_fr_expansion}
    \kappa^{FR}_{ij}=4-k_i -k_j +3 \triangle_{ij} + 2 \square_{ij} + \pentagon_{ij} \text{,}
\end{equation}
where $k_i$ denotes the degree of vertex $i$ and $\triangle_{ij}$, $\square_{ij}$ and $\pentagon_{ij}$ denote the number of triangles, squares and pentagons supported on edge $(ij)$, respectively \cite{tee}. This further simplifies for $k$-regular quasi-convex graphs,
\begin{equation}\label{eqn:fr_mean_field}
    \kappa^{FR}_{ij}= 4-2k + 3 \triangle_{ij} +2 \square_{ij} +\pentagon_{ij}  \text{.}
\end{equation}
These expressions show that, when degrees don't vary too much, combinatorial curvature is mainly determined by the number of local loops based on an edge, triangles contributing the most and pentagons the least. 

\subsection{Ollivier curvature}
There are several variants of Ollivier combinatorial curvature (OR) \cite{olli0, olli1, olli2, olli3, olli4}. Here we shall discuss only the simplest one, to be used to formulate general relativity on networks. Its basic idea is to reproduce in a discrete setting the geometric meaning of Ricci curvature.

While often introduced as a derivative expression of the metric, Ricci curvature on manifolds is a measure of how much infinitesimal spheres around a point contract (positive Ricci curvature) or expand (negative Ricci curvature) when they are parallel-transported along a geodesic with a given tangent vector at the point under consideration. The Ollivier curvature is simply a discrete version of the same measure. For two vertices $i$ and $j$ it compares the Wasserstein (or earth-mover) distance $W\left( \mu_i, \mu_j \right)$ between the two uniform probability measures $\mu_{i,j}$ on the spheres around $i$ and $j$ to the graph distance $d(i,j)$ and is defined as
\begin{equation}
\kappa^{OR}_{ij}= 1- {W\left( \mu_i, \mu_j \right) \over d(i,j)} \ .
\label{olli}
\end{equation}
The Wasserstein distance between two probability measures $\mu_i$ and $\mu_j$ on the graph is defined as
\begin{equation}
W\left( \mu_i, \mu_j \right) = {\rm inf} \sum_{i,j} \xi(i,j)d(i,j) \ ,
\label{wasser}
\end{equation}
where the infimum has to be taken over all couplings (or transference plans) $\xi(i,j)$ i.e. over all plans on how to transport a unit mass distributed according to $\mu_i$ on the unit sphere around $i$ to the same mass distributed according to $\mu_j$ on the unit sphere around $j$. 
\begin{equation}
\sum_j \xi (i,j) = \mu_i \ , \qquad 
\sum_i \xi (i,j) = \mu_j \ .
\label{transplan}
\end{equation}

While extremely intuitive from a geometric point of view, the OR seems even more cumbersome to compute in the generic case than the FR. This is why easier approximations have been developed \cite{klit1, klit2, klit3}. Fortunately, however, these are not really necessary since the OR also simplifies substantially on quasi-convex graphs satisfying the independent short cycles condition \cite{comb1}. In this case it takes the form \cite{comb2}, 
\begin{eqnarray}\label{equation:NewOllivCurv}
\kappa^{OR}_{ij}=\frac{\triangle_{ij}} {k_i\land k_j}-\left[1-\frac{1+\triangle_{ij} +\square_{ij}}{k_i\lor k_j}-\frac{1}{ k_i\land k_jj}\right]_+\nonumber\\
- \left(\frac{\triangle_{ij}}{k_i\land k_j}-\frac{\triangle_{ij}}{k_i\lor k_j}\right)\lor \left(1-\frac{1+\triangle_{ij} +\square_{ij} +\pentagon_{ij}}{k_i\lor k_j}-\frac{1}{k_i\land k_j}\right).
\end{eqnarray}
where the subscript ``+" is defined as $[\alpha]_+ = {\rm max} (0, \alpha)$ and  $\alpha\lor\beta:=\max(\alpha,\beta)$, $\alpha \land\beta:=\min(\alpha,\beta)$. For a $k$-regular quasi-convex graph this becomes
\begin{equation}
\kappa^{OR}_{ij}= \frac{\triangle_{ij}} {k}-\left[1-\frac{2+\triangle_{ij} +\square_{ij}}{k}\right]_+ - \left [1-\frac{2+\triangle_{ij} +\square_{ij} +\pentagon_{ij}}{k}\right]_+ \ .
\label{orcr}
\end{equation}

This shows that, when the degree distribution is rather peaked, also the OR of graphs is mostly determined by the number of short loops on an edge, where by short one means triangles, squares and pentagons. In this sense, the ``geometry" of graphs is a theory of loops. The quasi-convexity condition of independent short loops requires that such loops can touch along one edge but they cannot overlap more than this. It has thus a very physical interpretation as an incompressibility condition for loops (or graphs themselves), analogous to the hard-core condition for a gas of Bose particles \cite{comb1}. As the hard-core condition prevents the infinite compressibility catastrophe of a Bose gas of particles, the corresponding hard-core requirement for loops also stabilizes a gas of loops on graphs, as we show below. 

\begin{figure}[t!]
	\includegraphics[width=9cm]{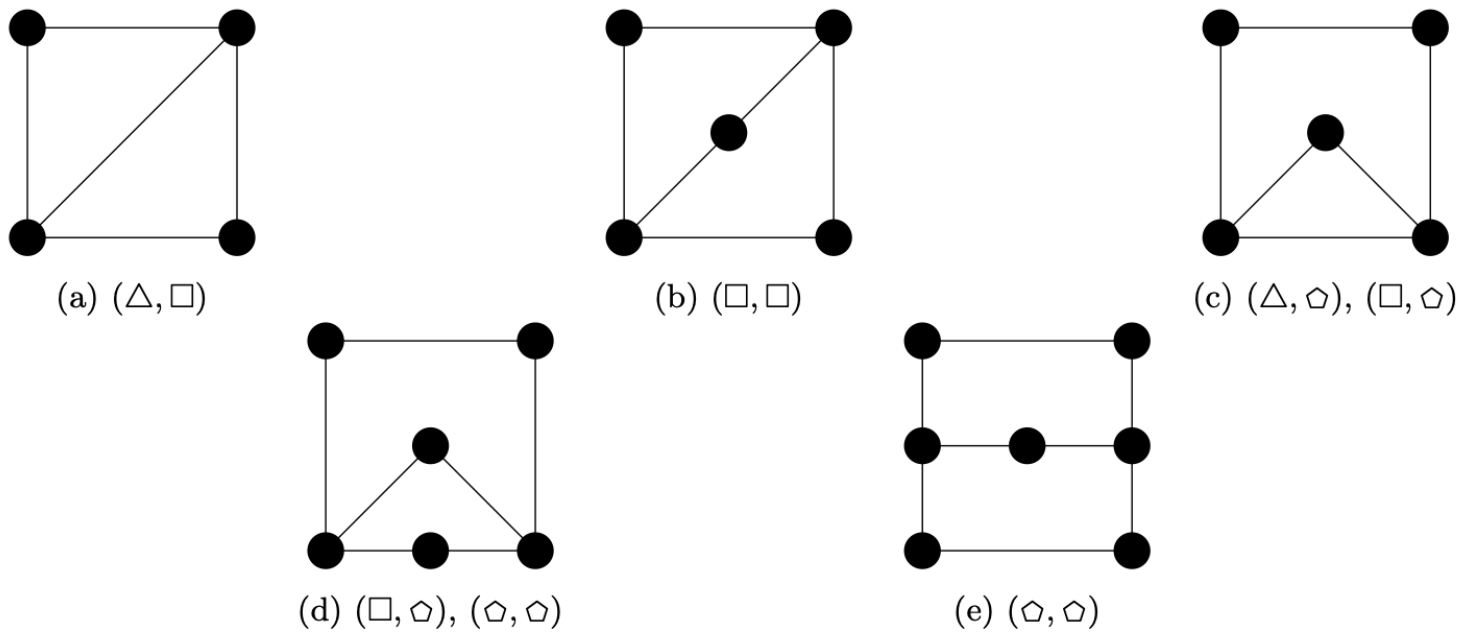}
	\vspace{-1cm}
	\caption{Excluded subgraph characterisation of the hard-core condition. Listed pairs for each subgraph give the types of
short cycles sharing more than one edge }
	\label{fig:Fig.1}
\end{figure}

The incompressibility condition can be also formulated as an excluded sub-graph condition \cite{comb2}. In Fig. \ref{fig:Fig.1} we list the subgraphs that must be absent for a graph to satisfy the independent short cycles condition and be thus quasi-convex. 

This leads immediately to another interpretation of quasi-convexity. On a Riemannian manifold, the metric component $g_{ij} (p)$ at a given point $p$ defines a unique infinitesimal surface element on the tangent space in the plane spanned by the two directions $i$ and $j$. The corresponding condition in the discrete graph setting requires that triangles, squares and pentagons, the fundamental local area elements, be uniquely defined by two edges. The excluded sub-graphs are exactly those for which this condition is not satisfied. In other words, quasi-convex graphs are those that satisfy the necessary condition for admitting a smooth continuum limit leading to a Riemannian manifold.

\subsection{Comparison of the Forman and Ollivier curvatures}
A general analytical comparison of the Forman and Ollivier curvatures is difficult because it is not simple to disentangle the contributions of the degrees and of the cycles. To make progress we will replace the degrees $k_i$ with their average value $\langle k\rangle$, assuming that the degree distribution is sufficiently peaked. We then represent the so obtained OR curvature as a ``mean field" term, defined as the expression in eq. \eqref{orcr} in which all brackets are simply summed up without taking into account the ``+" subscripts, plus a correction term. Essentially, the mean field term represents the dependence on the cycles alone, while the correction embodies deviations dependent on connectivity. This gives 
\begin{eqnarray}
    \langle k\rangle \kappa^{OR}_{ij} &&= \langle k\rangle \kappa^{ORMF}_{ij}+ \delta_{ij} \ ,
    \nonumber \\
    \langle k\rangle\kappa^{ORMF}_{ij} &&= 4-2\langle k\rangle +3 \triangle_{ij} + 2\square_{ij} +\pentagon_{ij} \ ,
    \label{mean}
\end{eqnarray}
where the correction to the mean field value $\kappa^{ORMF}_{ij}$ is given by
\begin{widetext}
\begin{equation}
\delta_{ij} = \begin{cases}
0 &\text{if $\langle k\rangle > 2+ \triangle_{ij}+\square_{ij}+\pentagon_{ij}$}\\
\langle k\rangle -2-\triangle_{ij}-\square_{ij}-\pentagon_{ij} &\text{if $2+\triangle_{ij}+\square_{ij} < \langle k\rangle \le 2+ \triangle_{ij}+\square_{ij} + \pentagon_{ij}$} \\
2\langle k\rangle -4-2\triangle_{ij}-2\square_{ij}-\pentagon_{ij} &\text{if $ \langle k\rangle  \le 2+\triangle_{ij}+\square_{ij}$}
\end{cases}
\label{mfcases}
\end{equation}
\end{widetext}

This shows that the properly enhanced Forman curvature coincides (up to an overall factor) with the mean field Ollivier Ricci curvature. 
\begin{equation}\label{eqn:mean_field_equivalence}
\langle k\rangle \kappa^{ORMF}_{ij} = \kappa^{FR}_{ij} \ .
\end{equation} 
This result is remarkable, given that the two discrete curvature constructions have completely different origins. It is also an indication that the combinatorial dependence on the number of cycles is unique for any local discrete curvature measure. 

The two curvatures become identical (up to an overall factor) when the quantity $I_{ij}$ defined by 
\begin{equation}
I_{ij} = \langle k \rangle -2 -\triangle_{ij} -\square_{ij} -\pentagon_{ij} \ ,
\label{lq}
\end{equation}
satisfies $I_{ij} \ge 0$, $\forall i,j$. When $I_{ij} >0$, we fall in the first case of eq. (\ref{mfcases}) and the correction term vanishes. When $I_{ij} = 0$, the second case of eq. (\ref{mfcases}) is realized and the correction term vanishes because it coincides with $I_{ij}$. The condition $I_{ij} \ge 0$, $\forall i,j$ is realized for graphs with large connectivity and sparse cycles or, e.g., for all regular and semi-regular planar and hyperbolic tessellations involving triangles, squares and pentagons only. For these cases the appropriately enhanced Forman curvature and the Ollivier curvature coincide (up to a factor). 

The case $I_{ij}=0$ is particularly interesting. This is the case for tilings of $D$-dimensional space, for which the skeleton degree $k = 2D$, so that at each vertex one can choose to ``move" forward or backwards along $D$ possible directions and where each of the edges supports exactly $k-2$ elementary faces, triangles, squares or pentagons . In this case the OR becomes
\begin{equation}
\kappa^{OR}_{ij} =  {\triangle_{ij} - \pentagon_{ij} \over k} \ .
\label{newor}
\end{equation}
This demonstrates that, in this geometric framework, triangles contribute positive curvature, squares contribute zero curvature, and pentagons contribute negative curvature. This is intuitively clear: triangles act as elementary units where ``geodesics" combinatorially converge, squares where they remain parallel, and pentagons where they diverge.

To further test the remarkable equivalence (\ref{eqn:mean_field_equivalence}) one can generate a variety of random graphs with varying connectivity and edge density and compute both curvatures numerically to compare them \cite{tee}. For this simulation we consider Erd{\"o}s-R{\'e}nyi (ER) random graphs (see, e.g., \cite{graphrev}) using a varying link probability $p$ in the range $0.01$ to $0.1$, after which we manually enforce the independent short-cycle condition by removing edges that violate it. For a fixed number $N=100$ of vertices this generates graphs with an average degree $\langle k\rangle >1$ above the critical threshold for the emergence of a large connected subgraph. As $p$ increases, the edge density of the graph increases, together with the density of short cycles. For each edge we compute $I_{ij}$ in (\ref{lq}), from which we build an average value: we expect that the two measures of curvature become closer and closer (up to the factor of $\langle k\rangle$) when this average of $I_{ij}$ approaches zero. 

\begin{figure}[ht]
	\centering
	\includegraphics[width=9cm]{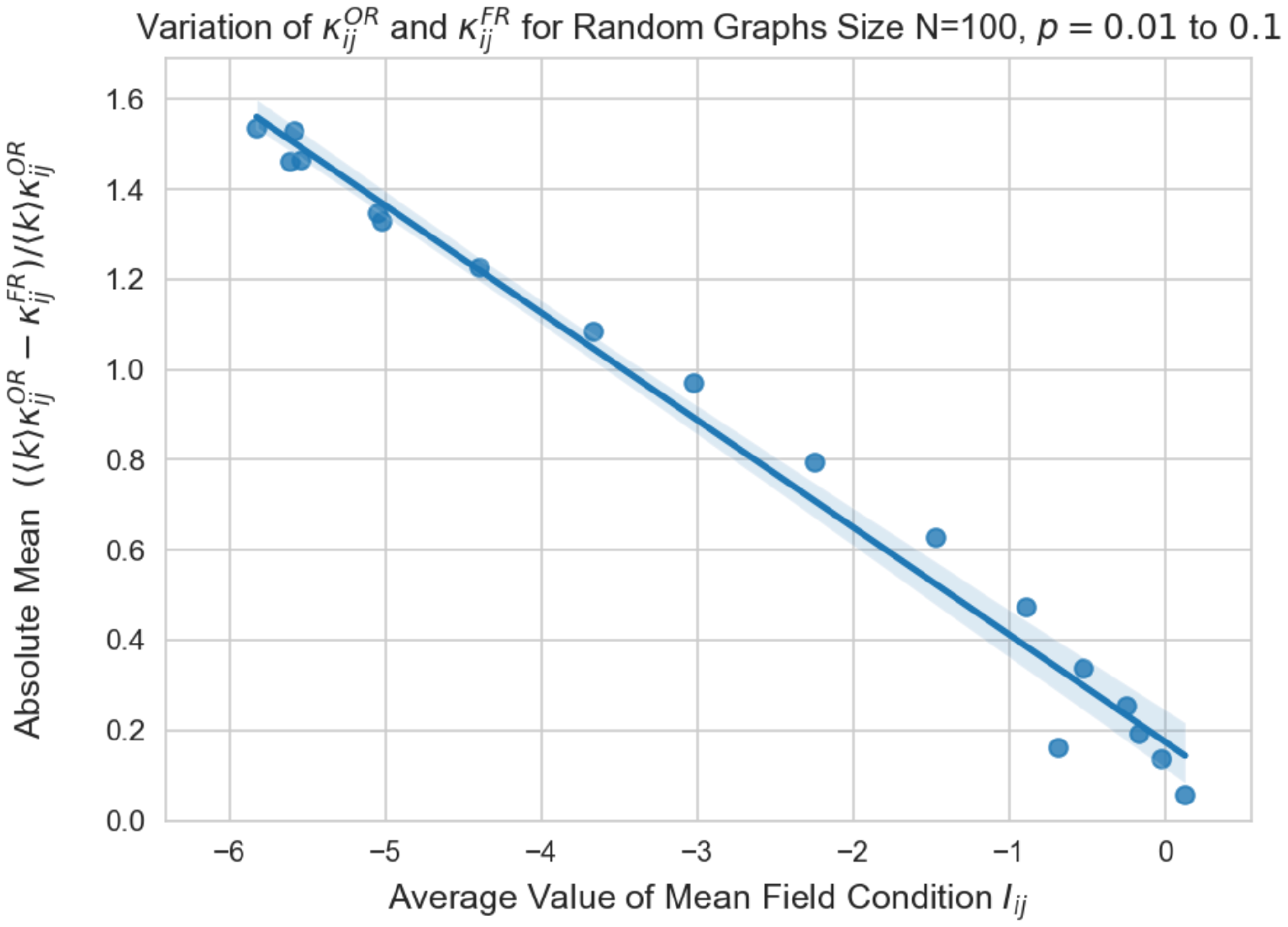}
	\caption{For $N=100$ vertices we generate random Erd{\"o}s-R{\'e}nyi graphs, varying the link probability $p$. For a collection of $10$ graphs for each value of $p$, we compute the average fractional difference between $\langle k\rangle \kappa^{OR}_{ij}$ and $\kappa^{FR}_{ij}$, which we plot against the average of $I_{ij}=\langle k\rangle-2-\triangle_{ij}-\square_{ij} -\pentagon_{ij}$. The blue dots represent the raw simulation values, with the blue line representing the linear regression fit, shading to cover the 95\% confidence interval for the regression.}
	\label{fig:Fig. 2}
\end{figure}

In \ref{fig:Fig. 2} we plot the average fractional difference between the two curvature measures, $(\langle k\rangle \kappa^{OR}_{ij} - \kappa^{FR}_{ij})/\langle k\rangle \kappa^{OR}_{ij}$ over all edges in randomly generated graphs against the average value of $I_{ij}$. For each link probability $p$ we generate $10$ graphs, to avoid the results being skewed by unusual graph configurations 
This simulation clearly shows that, here too, the two curvature measures differ when $I_{ij}<0$ but, as $I_{ij}$ increases and approaches zero they converge to the same value. 

\section{Convergence of Ollivier curvature to Ricci curvature}
To formulate general relativity on networks using a discrete curvature measure, it is essential to ensure that there is a regime in which this measure is mapped onto continuum Ricci curvature, which governs the large-scale Einstein equations. We will discuss two such limits. Here we show that Ollivier curvature converges to the continuum Ricci curvature as the network increasingly approximates a smooth manifold with a denser mesh \cite{vanHoorn}. Later we will show that even regular graphs, which are not dense, can define the 1-skeleton of a 2-cell complex which is a tessellation of hyperbolic space so that the geometric curvature is determined by the combinatorial one via the cycle structure at each vertex. 

When working with purely combinatorial discretizations of curvature, associating a network with a manifold and its Ricci curvature requires assigning a distance scale to the graph’s edges. In the context of emergent geometry from a statistical network model, the simplest approach is to assume a fixed edge length, as we will discuss below. If the manifold is already given, a more natural choice is to weight each edge by the corresponding geodesic distance on the manifold. This allows one to meaningfully consider the limit of vanishing Ollivier balls centered on points of the manifold.

A notable example of networks defined over manifolds is the class of random geometric graphs (RGG) \cite{gilbert} (for a review see \cite{penrose}), which are constructed as follows. Given a manifold $M$, one sprinkles it with a Poisson point process (PPP) of rate $r = O(n)$, which determines the average density of points on the manifold. Two points on the manifold are connected if and only if their manifold distance does not exceed a threshold $\delta_n$. The so obtained graph constitutes a random discretization of the manifold on a scale determined by $n$. This graph is the 1-skeleton of a Vietoris-Rips complex \cite{kahle}, whose topology was proven to converge to the manifold topology \cite{latsche}. In \cite{vanHoorn} it was shown that also its geometry converges to the manifold geometry. 

Consider a $D$-dimensional manifold$M$ and a point $x$ on it. Then pick a unit vector $v$ at $x$ and add a second point $y$ at a distance $\delta_n$ from $x$ along the geodesic in direction $v$. Finally let us consider a RGG on $M$ with rate $r=O(n)$ and connection threshold $\delta_n$ dependent on $n$. The graph distances are weighted by the manifold distances: for every pair of vertices $z$ and $w$ on the graph, the graph distance $d(z, w)$ is the sum of the manifold distances on the shortest graph path connecting $z$ and $w$. 

We want to compute the Ollivier curvature 
\begin{equation}
\kappa^{OR} (x, y) = 1-{W\left( \mu_x, \mu_y \right) \over \delta_n} \ ,
\label{oc1}
\end{equation}
as a function of $n$, where we consider uniform probability distributions $\mu$ on solid balls of radius $\delta_n$, instead of spheres,
\begin{eqnarray}
\mu_x (z) &&= {1\over |B\left( x,\delta_n\right)| }\ , \qquad  {\rm if}\ \ z\in B \left( x, \delta_n \right) \ ,
\nonumber \\
\mu_y (z) &&= {1\over |B\left( y,\delta_n\right)|} \ , \qquad  {\rm if}\ \ z\in B \left( y, \delta_n \right) \ ,
\label{balls}
\end{eqnarray}
with $B\left( x, \delta_n \right) = \{ z \in {\rm RGG}: d(x,z) \le \delta_n \}$ and a corresponding definition for the ball around $y$. The expression $|B|$ denotes the number of RGG vertices contained in the ball $B$, which, of course, depends on $n$. This is one the variants of the Ollivier curvature, which is the most appropriate for the convergence computation. 

With these settings one can show that 
\begin{equation}
{\rm lim}_{n\to \infty} {\langle \kappa^{OR} (x,y) \rangle \over \delta_n^2} = {\rm Ric(v,v) \over 2 (D+2)} \ ,
\label{oc2}
\end{equation}
if the ball radius shrinks to zero as,
\begin{eqnarray}
\delta_n &&= O \left( n^{-\alpha} \right) \ ,
\nonumber \\
\alpha &&< {1\over 3D} \ .
\label{oc3}
\end{eqnarray}
The expectation value $< \dots >$ is with respect to the RGG ensemble, in which $\kappa^{OR}$ is a random variable and Ric(v,v) denotes the continuum Ricci curvature at $x$ in direction $v$. The above result is strong, in the sense that it does not only imply the convergence of the Ollivier curvature to the Ricci curvature of the underlying manifold, but also the concentration of the random $\kappa^{OR}$ around its expected value, 
\begin{equation}
{\rm lim}_{n\to \infty} {\rm Prob} \left[ \left| {\langle \kappa^{OR} (x,y) \rangle \over \delta_n^2}-{\rm Ric(v,v) \over 2 (D+2)} \right| > \epsilon \right] = 0 \ ,
\label{oc4}
\end{equation}
for any $\epsilon > 0$. Since the expected average degree of the RGG behaves as 
\begin{equation}
\langle k \rangle = n {\rm vol} \left[ B_M \left( \cdot, \delta_n \right) \right] = O\left( n \delta_n^D \right) = O\left( n^{1-\alpha D} \right) \ ,
\label{oc5}
\end{equation}
with $B_M$ the ball in the manifold, the second condition in (\ref{oc3}) requires that the graph be dense or sparse but not ultrasparse (or truly sparse), in which case $\alpha = 1/D$. This shows that, for a sufficiently dense sampling of a Riemannian manifold $M$, a RRG $G$ will approximate $M$ as a metric space. 

To conclude this section on discrete curvature, we would like to stress that the discussed Forman and Ollivier curvatures are not the only notions of discrete curvature that have been proposed. In particular we would like to mention the Bakry-Emery curvature \cite{bakry} and the resistance curvature \cite{resistance}, although we will not discuss them in detail.

\section{The emergence of hyperbolic manifolds} 
General relativity is a theory of dynamical manifolds governed by the Einstein-Hilbert action. At the classical level, the stationary points of this action, given by the Einstein equations, determine the space-time geometry for a specified matter distribution, or the vacuum geometry in the absence of matter. At the quantum level, one expects the Euclidean partition function, formally defined as a sum over all manifolds, to capture the quantum fluctuations around the classical solution. Yet, this is only well-defined when the manifold is discretized; otherwise, the Euclidean Einstein-Hilbert action is unbounded from below \cite{cdt1}. Euclidean dynamical triangulations (EDT) still fail: they produce pathological manifolds with fractal dimensions. Introducing a preferred time foliation, as in causal dynamical triangulations, seems to cure the problem but, at least in the analytically tractable 2D case, it leads to relativistic invariance breaking at high energies, exactly as in Horawa-Lifshitz gravity \cite{cdthl}. 

Here we will dispense altogether with the notion of a manifold. Using the Ollivier curvature as a combinatorial analogue of the Ricci curvature entering the Euclidean Einstein-Hilbert action we shall formulate a gravitational partition function as a sum over networks. The central question is whether well-behaved, fluctuating geometric manifolds can emerge spontaneously from such a formulation. 

\subsection{Cycle condensation}
We consider thus a statistical mechanics model defined as a sum on quasi-convex 2D-regular graphs (QCG) with $N$ vertices. The partition function is 
\begin{equation}
{\cal Z} = \sum_{QCG} {\rm e}^{-{1\over g} H} \ ,
\label{comqg}
\end{equation} 
where $g$ is the dimensionless coupling and the Hamiltonian is defined as 
\begin{equation}
H = -{2D} \sum_{i \in G} \kappa (i) = -{2D}\sum_{i\in G} \sum_{j\sim i} \kappa (ij) \ ,
\label{deh}
\end{equation}
where we denote by $j \sim i$ the neighbour vertices $j$ to vertex $i$ in graph $G$, i.e. those connected to $i$ by one edge $(ij)$, $2D$ is the degree of the regular graph and $\kappa (ij)$ is the Ollivier curvature (\ref{olli}) of edge $(ij)$ (from now on we omit the superscript ``OR" since we will use only the Ollivier curvature). Note also that the constraints of quasi-convexity can alternatively be dynamically implemented by adding Kronecker-delta terms to the Hamiltonian \cite{evnin}.  I

It is important to stress that, contrary to the previously discussed case of random geometric graphs, there is no embedding manifold assumed here. As we will show, however, at small coupling $g$ the dominant graphs in the partition function (\ref{comqg}) {\it define} themselves a manifold, specifically a surface, as 1-skeletons of 2-cell complexes. For this the constraint of quasi-convexity is a crucial ingredient; it is the free energy minimization under this constraint that leads to the emergence of manifolds.  

Using the Ollivier curvature mean field plus correction representation (\ref{mean}) and (\ref{mfcases}), we can split the Hamiltonian into a global and a local contribution, 
\begin{eqnarray}
H &&= H_{\rm global} + H_{\rm local} \ ,
\nonumber \\
H_{\rm global} &&= 16 \left( {D(D-1)\over 2} N - {9\over 8} \triangle - \square- {5\over 8} \pentagon \right) \ ,
\nonumber \\
H_{\rm local} &&= \sum_{(ij) \in E_1} \left( (\triangle_{ij} + \square_{ij} ) - (2D-2) \right) 
\nonumber \\
&&+ \sum_{(ij) \in E_2} \left( (\triangle_{ij} + \square_{ij} +\pentagon_{ij}) - (2D-2) \right) \ ,
\label{globallocal}
\end{eqnarray}
where $\triangle$, $\square$ and $\pentagon$ are the total numbers of triangles, squares and pentagons on the graph and $E_1$ and $E_2$ are the ensembles of edges for which the respective summands are strictly positive. 

The global term in (\ref{globallocal}) corresponds (up to a constant factor) to the Hamiltonian derived from the Forman curvature. It represents the Hamiltonian of a matrix model that depends solely on the total number of cycles in the graph. The energy is maximized for tree-like random regular graphs, which contain no cycles, and decreases as short cycles form. 
Among these, triangles contribute the most significant energy reduction, while pentagons contribute the least. The quasi-convexity condition, however, implies that a triangle can share an edge only with a pentagon, not with another triangle or a square, see Fig. \ref{fig:Fig.1}. But it is energetically favourable to replace a triangle-pentagon couple with two squares, since $9/8 + 5/8 < 2$. Therefore, triangles and pentagons can survive only as isolated defects but are excluded for an homogenous ground state and we can henceforth neglect them. The global Hamiltonian is then minimized by increasing the number of squares. The local Hamiltonian, however penalizes edges for which the number of squares is higher than (2D-2). The total expression (\ref{orcr}) for the Ollivier curvature (with no triangles and pentagons) shows that these two effects cancel exactly when a configuration with D(D-1)/2 squares per vertex and (2D-2) squares per edge is reached. Configurations with a higher number of squares (the maximum per edge is (2D-1) by regularity) are degenerate with this one. What happens is that, decreasing the coupling from the random configuration with sparse cycles, the number of squares at an edge increases until there are exactly (2D-2) squares based there. After this it pays energetically to increase the number of squares at another edge. This can continue until there are (2D-2) squares at each edge and ND(D-1)/2 globally. This is a hypercubic complex, which identifies D as the topological dimension of the ground state. For D=2, this is a torus graph. The stability of the torus graph ground state has been discussed in detail and analyzed numerically in \cite{comb2}. To make numerical simulations less computationally intensive one can use bipartite graphs, for which there are no odd-length cycles. From here onward, we focus on the case $D=2$ for which, as we now show, the ground state encodes the degrees of freedom of a two-dimensional (2D) holographic screen, analogous to the framework proposed in Verlinde’s entropic gravity model \cite{verlinde1, verlinde2}. 

\begin{figure}[t!]
	\includegraphics[width=10cm]{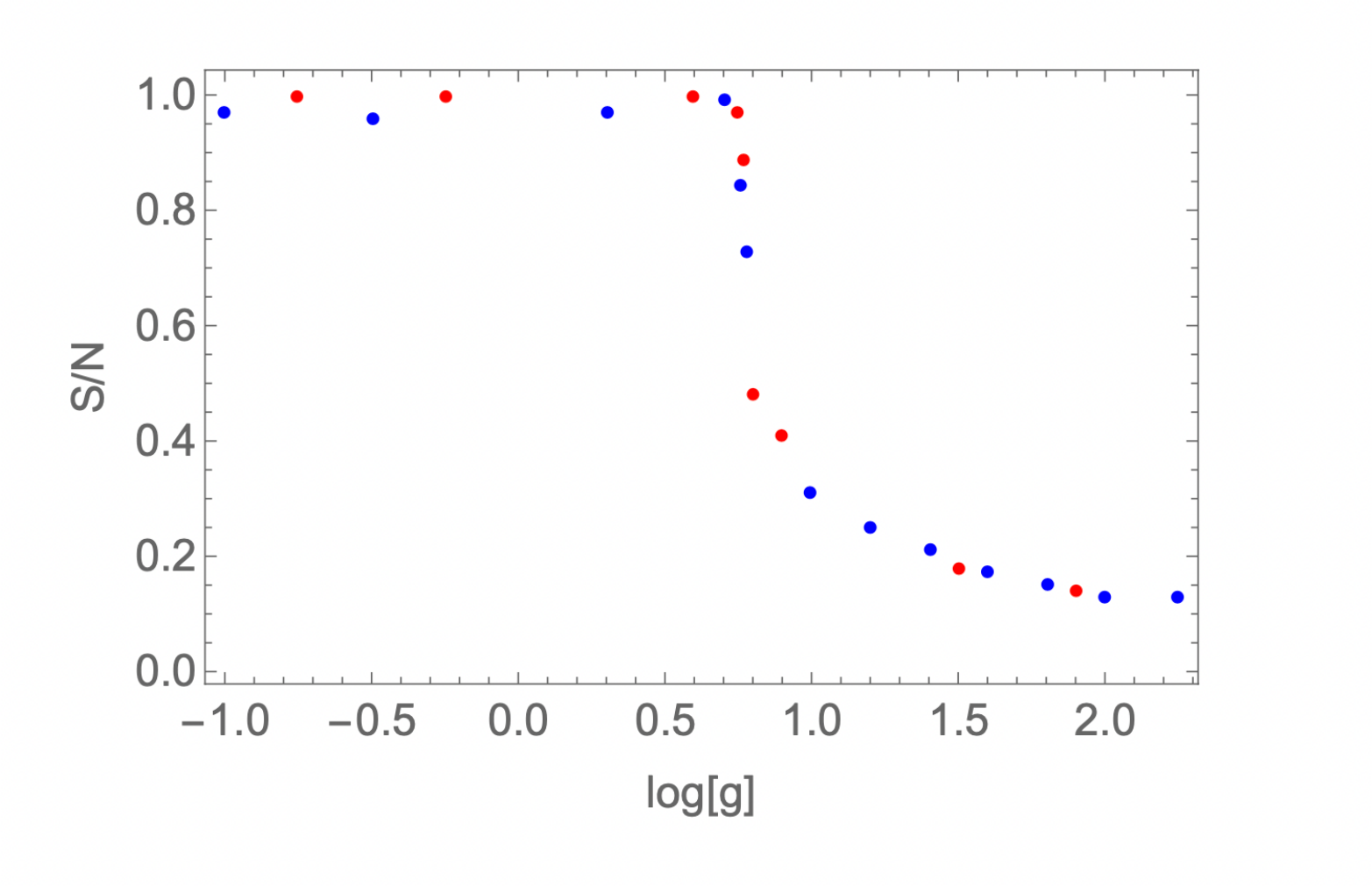}
	\vspace{-0.3cm}
	\caption{The order parameter $S/N$ for the random-to-geometric phase transition for $D=2$. }
	\label{fig:Fig.3}
\end{figure}

If only the global term is retained, the model undergoes a first-order phase transition in which the graph decomposes into isolated, weakly interacting hypercubic complexes \cite{comb2, gorsky1, gorsky2}. If the full Hamiltonian is used, instead the model undergoes a continuous phase transition from a random graph phase to a geometric phase in which graphs become tessellations of smooth surfaces \cite{comb1, comb2, eff1, eff2, universe}. This phase transition is due to a condensation of square loops on the graph, while triangles and pentagons are suppressed and survive at best as isolated defects \cite{comb2}. This is in agreement with a previous observation that the emergence of geometry in random networks is associated with clustering \cite{krioukov}. Random regular graphs are locally tree-like, with very sparse cycles, governed by a Poisson distribution with mean $(2D-1)^k/k$ for cycles of length $k$ \cite{oneil}. The order parameter for the transition is the density of squares $S/N$, which increases from $(3^4/4)/160= 0.126$ (for $N=160$ in the example below) in the random phase to reach the maximum value 1 for a torus graph. It is shown in Fig. \ref{fig:Fig.3} as a function of ${\rm log} (g)$. 

The red dots are obtained by starting with a torus graph and gradually increasing the coupling $g$; the blue dots, instead, correspond to decreasing the coupling from the random phase. Note the absence of hysteresis, as expected for a continuous phase transition.  As could be expected, the Monte Carlo algorithm does not always manage to reach an exact ground state with $S/N=1$ deep in the geometric phase when starting from a random graph configuration. Of course, several other diagnostics for a second order transition have been tested \cite{comb2}, with positive results.

\begin{figure}[t!]
	\includegraphics[width=9cm]{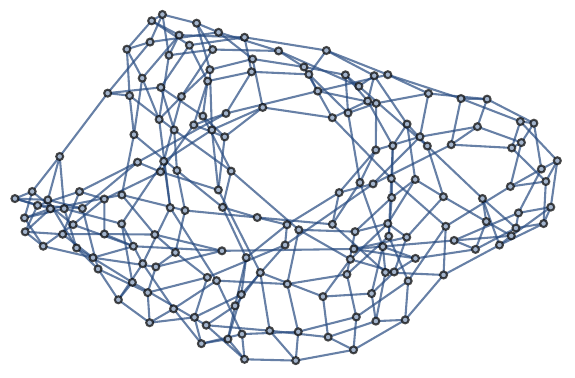}
	\vspace{-0.3cm}
	\caption{A two-torus at a small but finite value of the coupling $g$ in the geometric phase }
	\label{fig:Fig.4}
\end{figure}

Any finite graph has a genus, which is defined as the smallest genus of a surface on which it can be embedded as a 2-cell complex without any edge crossings \cite{genus}. Random graphs of $N$ vertices can be embedded without edge crossings on a Riemann surface of genus N \cite{expander}. Therefore, decreasing the coupling from its critical value amounts to  decreasing the genus of the emerged manifold until the ground state torus graph at $g=0$ is reached. In Fig. \ref{fig:Fig.4} we show the example of a two-torus state at a small but finite value of $g$ in the geometric phase.

\begin{figure}[t!]
	\includegraphics[width=8cm]{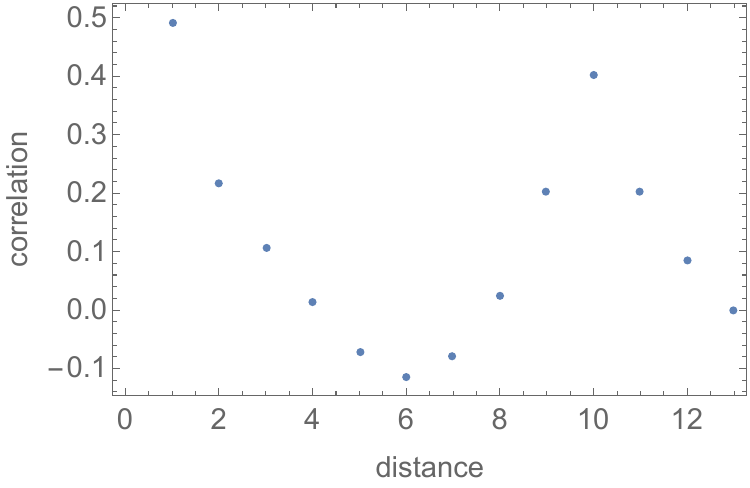}
	\vspace{-0.3cm}
	\caption{The typical behaviour of the connected correlation function between the numbers of squares at graph distance $d$ in the geometric phase. }
	\label{fig:Fig.5}
\end{figure}

As always in the case of continuous phase transitions, there in an emergent scale. This is the correlation length $\xi(g)$, which vanishes in the ground state at $g=0$ and diverges at the critical point. In the present case, let us consider the connected correlation function
\begin{equation}
c (d, g) = {\langle \left( \square_i- \bar \square \right ) \left( \square_j- \bar \square \right ) \rangle \over \sigma^2 (\square)} \ ,
\label{correlation}
\end{equation}
between the number of squares $\square_i$ and $\square_j$ based on vertices $i$ and $j$ at fixed graph distance $d$, where $\sigma^2 (\square)$ denotes the variance (over the vertices) of the number of squares. The typical behaviour of this correlation in the geometric phase, as a function of $d$ (at fixed $g$) is shown in Fig. \ref{fig:Fig.5}.

At extremely small distances the correlation is high since, given a square based at vertex $i$, there are always at least two vertices at distance 1 and one vertex at distance 2 that also support a square. At intermediate distances the correlation turns negative, because there is no correlation between squares on these distances but we are in the geometric phase, in which the average square number $\bar \square$ is large. On these intermediate distances the graph behaves like a random graph. Correlations then suddenly pick up at the distance $\xi(g) =10$, before decaying again to zero on even larger distances. In this case, the approach to zero is not due to absence of correlations but, rather, to each vertex supporting the same number of squares $\bar \square$. This is the geometric manifold phase on large scales. As usual, the correlation length $\xi(g)$ separates the ordered phase (here the geometric phase), realized at large scales, from the remnants of the disordered phase (here the random phase) at small scales. The graphs in the geometric phase behave thus as geometric manifolds on distances larger than $\xi(g)$ and as random graphs below this scale. To make contact with manifolds, we shall henceforth endow the links of the graph with a uniform length scale $\ell$ and identify the correlation length with the Planck length, $\xi(g) \ell = \ell_{\rm P}(g)$. With this definition, the Planck length vanishes at $g=0$ whereas at $g=g_{\rm cr}$ it equals the graph diameter. 

Note that, strictly speaking, the continuous phase transition in this model of emergent geometry is fully characterized not just by the mean number of squares, but by the entire distribution of squares across the ensemble. In the random phase, this distribution is a Poisson distribution, with variance equal to its mean, as previously noted. In the geometric phase, the distribution becomes sharply peaked around its mean, approaching a delta function in the limit of zero coupling. Both phases are governed by a finite correlation length, which sets the scale for fluctuations and keeps the variance finite. At the critical point, however, the distribution becomes scale-free, and the variance diverges. This signals the emergence of ensemble fluctuations that include clusters of 4-cycles and random links on all length scales.

Two-dimensional quantum gravity is often described as both trivial and pathological. It is considered trivial because it lacks local degrees of freedom, only the topological genus of the manifold plays a role. At the same time, it is deemed pathological because large quantum fluctuations cause the geometry to crumple into fractal structures with a Hausdorff dimension 
$d_{\rm H}= 4$. This behaviour is observed in both the continuum approach, such as Liouville theory, and in discrete formulations like matrix models or Euclidean dynamical triangulations (see, e.g., \cite{seiberg, tasi} for reviews). In contrast, the model studied here belongs to a different universality class. It features two distinct phases, one of which does not admit a manifold interpretation: this ensures the existence of local degrees of freedom at scales below the Planck length. Meanwhile, in the geometric phase, the quasi-convex graph structure gives rise to smooth manifolds with a Hausdorff dimension $d_{\rm H}= 2$. 

\begin{figure}[t!]
	\includegraphics[width=8cm]{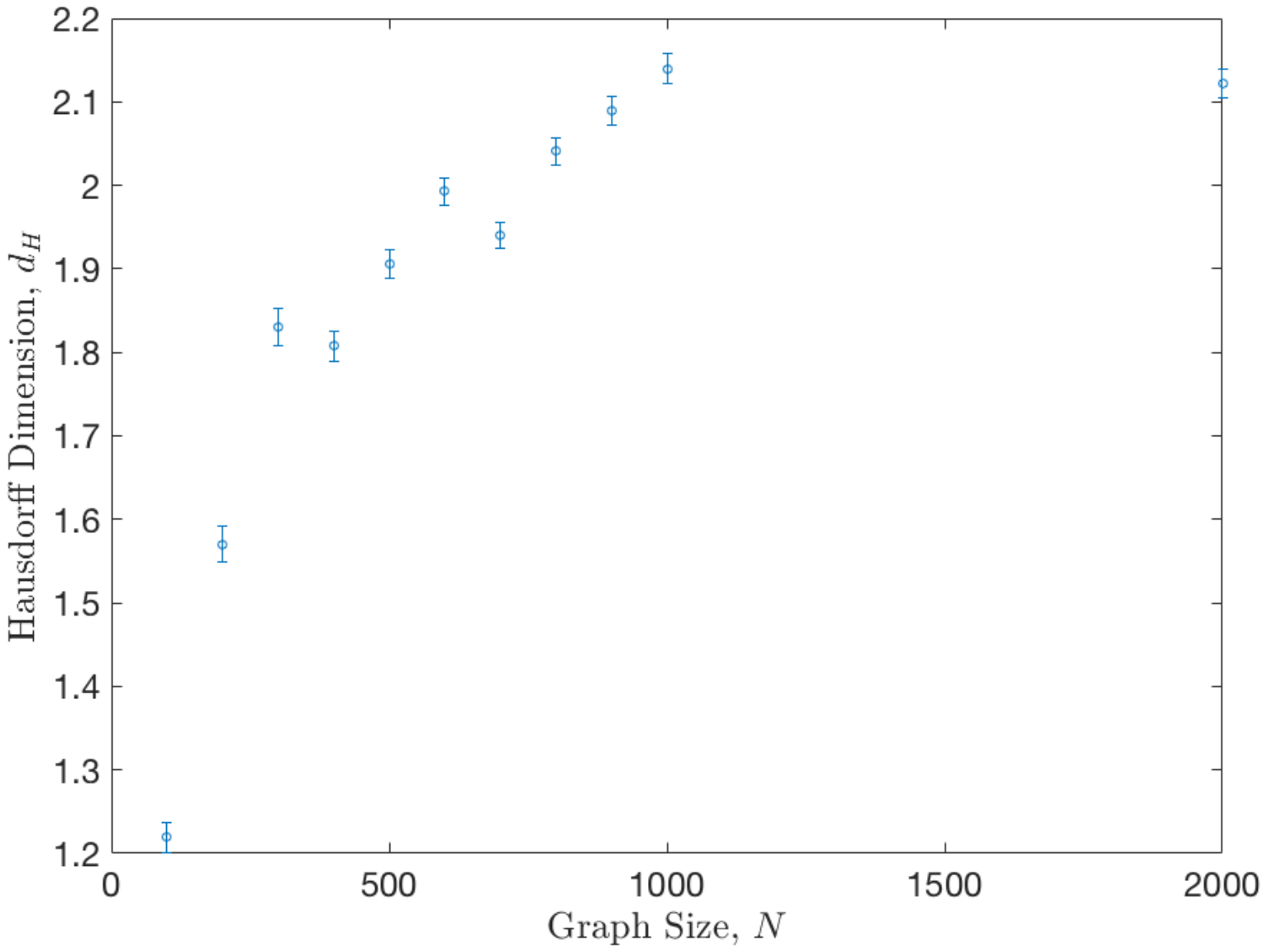}
	\vspace{-0.3cm}
	\caption{The Hausdorff dimension deep in the geometric phase as a function of size $N$.  }
	\label{fig:Fig.6}
\end{figure}

In this context, the Hausdorff dimension characterizes how the graph’s diameter ${\cal D}$ scales with the number $N$ of vertices 
\begin{equation}
\mathcal{D} \sim N^{1/d_{\rm H}} \ .
\label{haus}
\end{equation}
This scaling relation captures the fractal nature of the emergent geometry in the large-scale limit. In the random phase, with logarithmic distance scaling, the Hausdorff dimension is infinite. Figure \ref{fig:Fig.6} shows the behaviour of $d_{\rm H}$ as a function of graph size $N$ deep within the geometric phase. For graph sizes up to $N=1000$, the Hausdorff dimension increases to values between 2.1 and 2.2. However, for significantly larger graphs, this growth is not only halted, it gradually reverses, with $d_{\rm H}$ approaching 2 (the gap in data between $N=1000$ and $N=2000$ reflects the high computational cost: simulations at each value of $N$ require months of CPU time). These findings support the conclusion that sufficiently large graphs in the geometric phase effectively approximate a smooth 2D manifold, unlike all previously studied models. This 2D manifold at large scales plays the role of a holographic screen analogous to what proposed in the framework of Verlinde's model \cite{verlinde1, verlinde2}, with some notable differences to be discussed below. 

\subsection{Geometry from combinatorics}
Regular graphs are ultrasparse and, as such, their combinatorial curvature is not sufficient to sample the continuum Ricci curvature on a manifold on which they are embedded, as we have shown above. There is, however, another way to map the combinatorial curvature onto a geometric curvature, as we now show. 

We will henceforth consider the limit $N \to \infty$, in which the emerging manifolds are the covering spaces of genus $g$ Riemann surfaces, i.e. infinite surfaces of negative curvature, so-called Cartan-Hadamard surfaces (for a review see \cite{shiga, anderson}). To simplify things, we will initially forget about the small scale randomness and, also, we will consider first configurations in which each vertex supports the same number $\square$ of squares. These correspond to a quantized constant negative curvature, as we now show. 

In the geometric phase, graphs become the $1$-skeleton of a $2$-cell embedding, for which the graph cycles are homeomorphic to open disks on a surface. The $2$-cell embedding is called a ``map”. A combinatorial map can be geometrized by assigning the same fixed geodesic length $\ell$ to each edge so that it becomes a tessellation of the surface, in which the combinatorial cycles become regular polygonal faces of the tessellation \cite{datta, maiti1, maiti2}.

Each edge of the map supports exactly two regular polygons, each vertex supports exactly four polygons. Let $k_i$, $i=1\dots 4$, be the number of edges of the 4 polygons surrounding a vertex and let us call the vector ${\bf k} = (k_1, k_2, k_3, k_4)$ the vertex type. For simplicity, we shall consider homogeneous maps, for which the edge type is the same for all vertices, so that the geometrization gives rise to semi-regular tessellations. For such maps, the tessellated surface is a constant curvature surface given by the angle sum parameter 
\begin{equation}
\alpha = \sum_{i=1}^4 {k_i-2\over k_i} \ .
\label{angsum}
\end{equation}
If $\alpha <2$ the surface is spherical, i.e. of positive constant curvature, if $\alpha=2$ it is the flat Euclidean plane and, finally, if $\alpha >2$ it is hyperbolic, i.e. of negative constant curvature \cite{datta, maiti1, maiti2}. 

For every hyperbolic tessellation there is exactly one geodesic length so that the sum of interior angles at each vertex sums exactly to $2\pi$. To see this, let us use the cosine rule for a hyperbolic $n$-gon,
\begin{equation}
{\rm sin} \left( {\theta \over 2} \right) = {{\rm cos} \left( {\pi \over n} \right) \over {\rm cosh} \left( {\ell \over 2R} \right)} \ ,
\label{hypercos}
\end{equation}
to write the total interior angle at a vertex of a tessellation as
\begin{equation} 
\sum_{i=1}^4 2 \ {\rm arcsin} \left( {{\rm cos} \left( {\pi \over k_i } \right)  \over {\rm cosh} \left( {\ell \over 2R} \right)} \right) = 2\pi \ ,
\label{intan}
\end{equation}
where $R$ is the radius of curvature of the Poincar\'e disk on which the tessellation is constructed. First of all let us note that this is a monotonically decreasing function of $\ell / R$. Second, for $\ell / R \to \infty$ the total interior angle vanishes. In the opposite limit $\ell / R \to 0$, the solution of (\ref{intan}) requires $\alpha=2$, i.e. the geometry becomes Euclidean. Therefore, given a specific semi-regular hyperbolic tesselation, there is exactly one possible parameter $\ell/R$ that solves (\ref{intan}). If the length unit $\ell $ is held fixed, the hyperbolic radius $R$ of the Poincar\'e disk varies and, with it, the curvature $K=-1/R^2$. Otherwise, we can describe varying curvature at a fixed radius by letting the length unit $\ell$ change. The tessellation realized by the graph minimizing the free energy at every coupling $g$ determines the value of the curvature $K(g)$ in units of $1/\ell^2$. 

\begin{figure}[t!]
	\includegraphics[width=8cm]{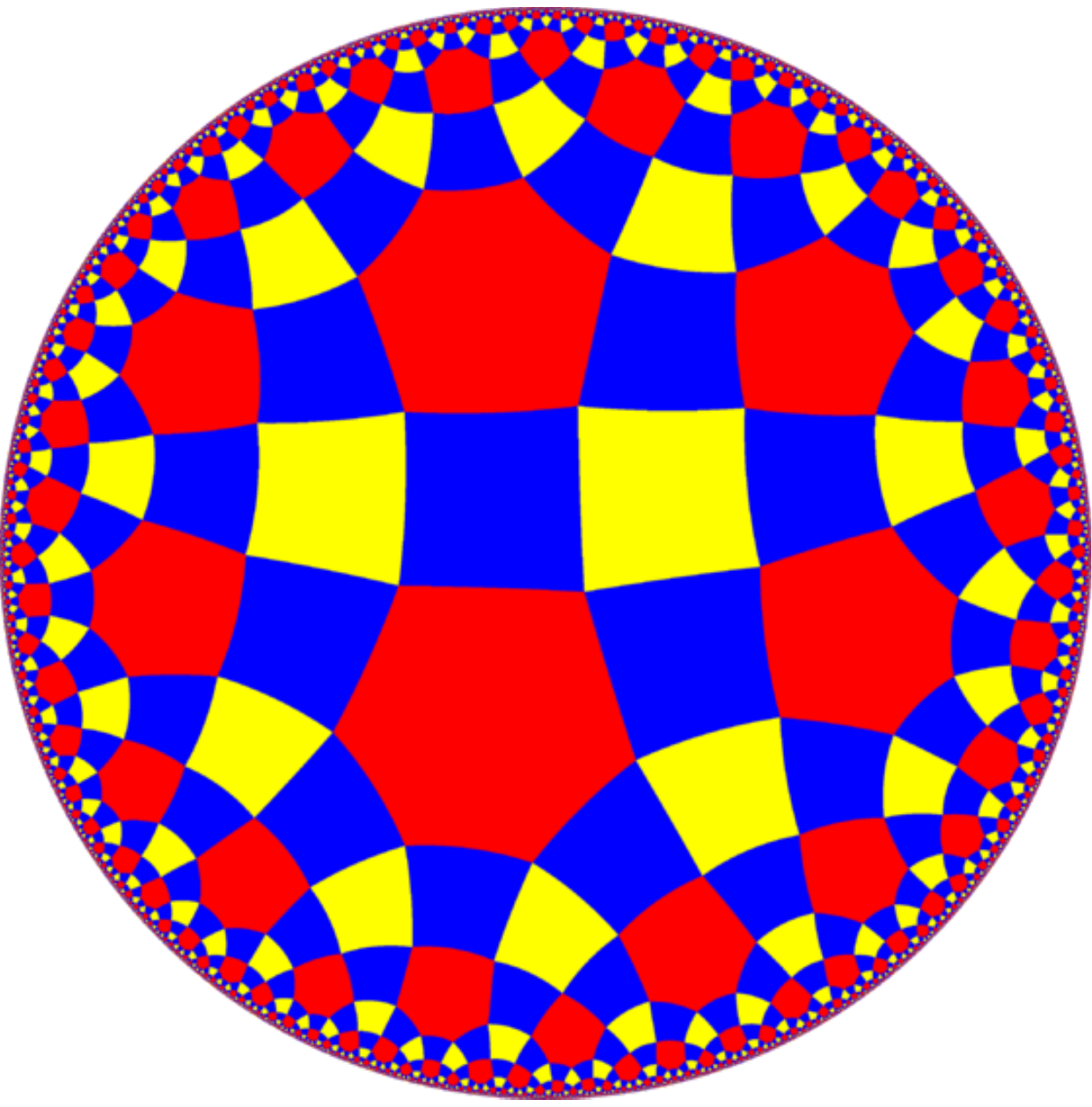}
	\vspace{-0.3cm}
	\caption{The 2D tessellation geometrizing the three-square-per-vertex infinite graph via its $1$-skeleton.}
	\label{fig:Fig.7}
\end{figure}

The geometric phase is realized by the condensation of squares and corresponds, thus to vertex types with at least one $k_i=4$ and all other $k_i= 4$ or $k_i \ge 6$ for all vertices. At each coupling, one of the four possible vertex types of the semi-regular tessellation, with one, two, three or four squares will characterize the minimum of the free energy, the remaining larger cycles determined by the detailed balance between the combinatorial energy and entropy. When the coupling is decreased from the critical value, the number of squares corresponding to the free-energy minimum will typically increase. For sufficiently large steps, one of the $k_i$ decreases from a higher value to 4. In order to maintain the total interior angle sum as $2\pi$, as in (\ref{intan}), the quantity $\ell /R$ has to decrease, which means the absolute value of the curvature of the tessellated surface typically decreases, until it vanishes for four squares around each vertex. In Figure \ref{fig:Fig.7}. we show one example of such a tessellation with three squares and one hexagon per vertex, corresponding to a constant negative-curvature surface at a small but non-zero value of the coupling $g$. 

Note that one cannot take the length scale $\ell $ to zero without the curvature also vanishing. In the infrared, the model is pushed to the completely ordered zero-coupling state corresponding to a flat geometry. At small, but non-vanishing coupling $g$ the surfaces emerging from the network gravity model are characterized by two connected scales, a fundamental length scale $\ell_{\rm P}$ and a radius of curvature $R$. At zero coupling the former vanishes and the latter diverges, at the critical coupling the former diverges and the latter vanishes. 

Via this construction, the purely combinatorial Ollivier curvature is mapped to geometric curvature. The value of this curvature is determined by the statistical balance between the combinatorial energy dependent only on square cycles and the entropy representing the possible choices of residual larger cycles. Of course, in the general case, the tessellation will not be semi-regular, corresponding to surfaces with possibly varying negative curvature, so-called Cartan-Hadamard surfaces \cite{shiga, anderson}. And finally, on scales smaller than $\ell_{\rm P}$, there are random microscopic degrees of freedom.

\section{Dynamics}
\begin{figure}[t!]
	\includegraphics[width=10cm]{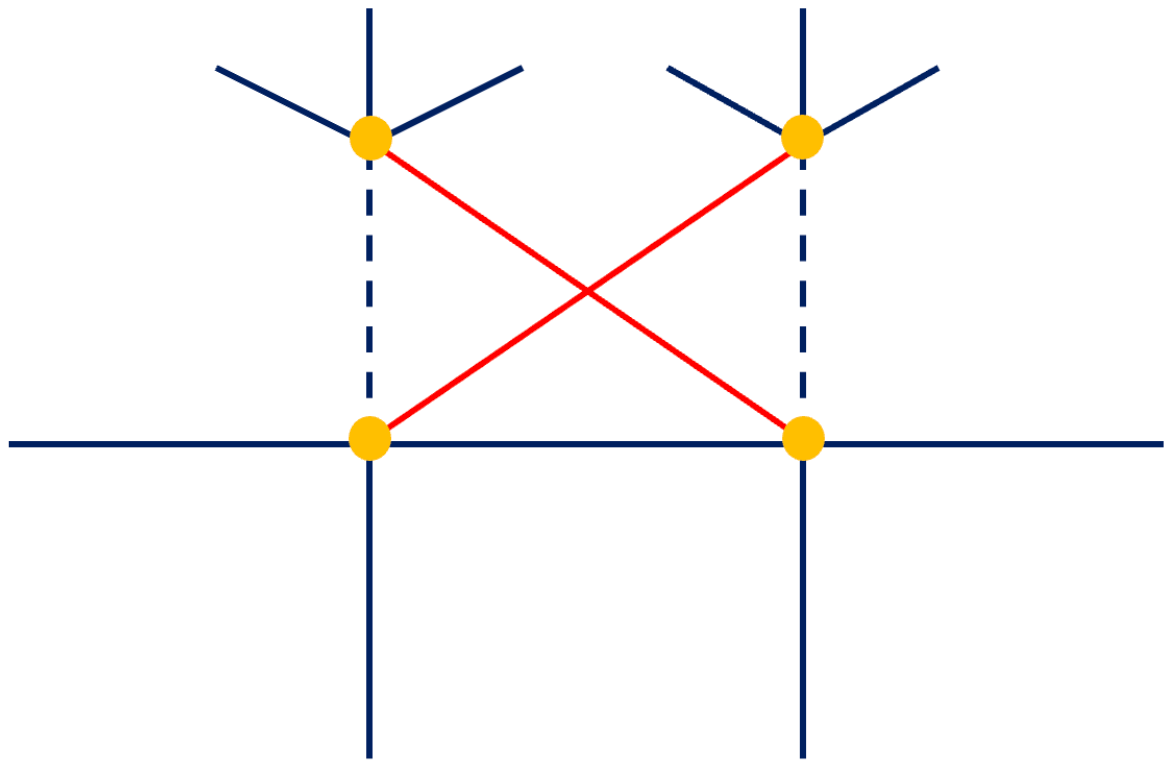}
	\vspace{-0.3cm}
	\caption{The regularity-preserving neighbourhood swap move.}
	\label{fig:Fig.8}
\end{figure}

Until now we have dealt with static configurations minimizing the free energy at a given coupling $g$. To analyze the dynamics, i.e. the fluctuations around these free energy minima, we need to introduce a universal Newtonian time $t$. Given an edge configuration $\{ e(t) \}$, a transition to a modified edge configuration $\{ e(t+1) \}$ takes place with a probability
\begin{eqnarray}
p &&= {1\over 1+ {\rm exp}({\Delta H/ g})} \ ,
\nonumber \\
\Delta H &&= H\left( \{ e(t+1) \} \right) - H\left( \{ e(t) \} \right) \ .
\label{glauber}
\end{eqnarray}
Of course, the transition has to be kinematically allowed, i.e. respect the regularity of the network and quasi-convexity and therefore must involve at least two edges. One such local, elementary fluctuation preserving regularity is the ``neighbourhood swap", shown in  Figure \ref{fig:Fig.8}. On a square lattice, e.g., this move replaces three squares with two pentagons. It represents thus a local curvature fluctuation which, by a simple combination of four-edge moves, can propagate on the lattice. 

Let us now consider a cluster of random phase at a small but non-vanishing value of $g$. The boundary of such a cluster is made by all vertices which support at least one non-accidental square, meaning that at least one of its neighbours supports more than one square. In general there are two types of fluctuations, one that changes shape and size of the clusters (akin to Glauber fluctuations in the Ising model) and one that corresponds to translations of the clusters at fixed order parameter (akin to Kawasaki fluctuations in the Ising model). Indeed, the energy of a cluster depends exclusively on the number $n$ of vertices in its bulk (supporting no short cycles) and on its boundary configuration. 
Therefore, a configuration replicating exactly the same cluster at a different location on the tessellated surface has exactly the same energy. A collective move implying a translation of the cluster costs no energy. Because of detailed balance in the equilibrium state, such zero modes will perform diffusive Brownian motion. Of course, it is an open question, beyond the scope of this review, how the collective translation can occur in terms of elementary edge moves. In any case the average velocity $v$ of this collective Brownian motion is limited by above by a velocity $c$ determined by the length $\ell$ and the inverse number of elementary moves per unit time, which are the two sole fundamental units in the model.

\subsection{Hyperbolic holography}
To proceed we shall consider a large-scale hydrodynamic approach consisting of point defects in Brownian motion on a constant negative-curvature surface. It turns out that curvature has a profound effect on Brownian motion, with momentous consequences: in the case of negative curvature it is anomalous and becomes asymptotically ballistic. 

Let us consider a constant negative curvature surface as the upper sheet of a two-sheeted hyperboloid embedded in 3D Minkwoski space with metric $(+1,+1, -1)$ as
\begin{equation}
{\bf x} = \begin{pmatrix} {1\over J} {\rm sinh} Jz \ {\rm cos} \theta \\ {1\over J} {\rm sinh} Jz \ {\rm sin} \theta \\ {1\over J} {\rm cosh} Jz \end{pmatrix} \ ,
\label{twosheets}
\end{equation}
The metric is $ds^2 = dz^2 + G^2(z) d\theta^2 $, with $G(z) = {{\rm sinh} Jz / J}$
with $J=1/R$ and $K=-J^2=-1/R^2$ the curvature. 

For $t{\cal D}J^2 \gg 1$ (with ${\cal D}$ the diffusion coefficient), the Brownian motion $b(t)$ on such a manifold can be expressed as 
\begin{equation}
b(t) = {\rm lim}_{T\to \infty} \left( {t\over T} b(T) + \delta b (t,T) \right) \ ,
\label{bridge}
\end{equation} 
where $\delta b(t,T)$ is the Brownian bridge, the Brownian motion $b(t)$ constrained to come back to the origin $x=o$ for $t = T$. The quantity ${\rm lim}_{T\to \infty} \delta x(t, T)$ is called the infinite Brownian loop \cite{anker}. If $b(t) = o(t)$ for $t\to \infty$, as in the typical case, the infinite Brownian loop is the Brownian motion itself. Not so, however, if the Brownian motion is ballistic, $b(t) = O(t)$ for $t\to \infty$. This is exactly the case in negative curvature. 

Brownian motion on a negative curvature surface decomposes in two independent radial and angular Brownian motions $(z_t, \theta_t)$ \cite{prat, kendall, hsu} (for a review see \cite{hsubrief, hsubook, arnaudon}). For $Jz \gg 1$, the angular component converges to a limiting angle $\theta_\infty$, while the radial component becomes ballistic, 
\begin{equation}
z_t= v t
\label{rad}
\end{equation}
where $v=J{\cal D}/2$ is the inertial velocity of the defect and ${\cal D}$ is the diffusion coefficient, which depends on its size. Asymptotically, the manifold coordinate $z$ is dynamically soldered to the absolute time and Brownian motion acquires a deterministic component describing inertia in negative curvature. The exponential volume growth due to negative curvature creates a surface of last scattering: after the typical time scale $1/vJ$ background fluctuations decouple and the defect moves on radially with the velocity $v$ it had after the last scattering \cite{eff1, eff2}. 

Let us focus first on this deterministic component. The asymptotic inertial trajectories (\ref{rad}) at different angles $\theta_{\infty}$ diverge because the negative curvature implies an exponential volume growth when $z$ increases. Up to an exponentially small correction, these trajectories are the same as the comoving geodesics on a 2D Lorentzian manifold of positive curvature $K = +J^2$, realized by the embedding 
\begin{equation}
{\bf x} = \begin{pmatrix} {1\over J} {\rm cosh} Jz \ {\rm cos} \theta \\ {1\over J} {\rm cosh} Jz \ {\rm sin} \theta \\ {1\over J} {\rm sinh} Jz \end{pmatrix} \ ,
\label{onesheet}
\end{equation}
with the metric
\begin{equation}
ds^2 = -dz^2 + {{\rm cosh^2}Jz\over J^2} d\theta^2 \ .
\label{metricone}
\end{equation}
This is 2D de Sitter space-time (for a review see \cite{strominger}), in which comoving geodesics diverge dynamically because of the Hubble expansion with parameter $J$ (see \cite{strominger}). Asymptotically, inertially diffusing observers in Riemannian negative curvature are equivalent, up to an exponentially small correction, to comoving observers in Lorentzian positive curvature \cite{eff1, eff2}. One can use both descriptions; however, with positive curvature, the universal time $t$ becomes the coordinate time of a Lorentzian manifold. This is a form of Wick equivalence arising between the two signs of curvature. 

In general, diffusion processes probe the intrinsic geometry of a manifold via the return probability kernel $K(t) =  {\rm Tr} \ {\rm exp}({\cal D} t \Delta)$. The quantity $d_{\rm s} = -2 d{\rm ln}\ K(t)/d{\rm ln}\ t$ is called the spectral dimension and measures the effective number of dimensions available to a random walker on the scale reached by time $t$, essentially on small scales for $t\to 0$ and on very large scales for $t \to \infty$. The spectral dimension is of paramount importance, since it is the dimension perceived by local physical processes. 

Unfortunately, this does not work as usual in the present framework, because of the dominant ballistic component of Brownian motion, which is much too fast to probe the local geometry. The Laplacian on a constant negative curvature manifold has a spectral gap
\begin{equation}
\lambda_0 = - {\rm lim}_{t\to \infty} {{\rm ln}\ K(t) \over t} = {(D-1)^2J^2\over 4}  \ ,
\label{spegap}
\end{equation} 
representing the bottom of the spectrum of the positive operator $-\Delta$, which is associated with asymptotic ballistic behaviour.  As a consequence, the return probabilities $K(t)$ are dominated by an exponential behaviour at large $t$  \cite{davies, grigorian},
\begin{equation}
K(t) \asymp {\left( 1+vJ t \right)^{(D-3)/2} \over (vJ t)^{D/2}} {\rm e}^{ -{(D-1)^2 \over 4} vJ  t} \ ,
\label{returnCH}
\end{equation}
which gives a spectral dimension linearly diverging in time. 

The large-scale local geometry, however, is probed by the sub-dominant random component of Brownian motion, the infinite Brownian loop \cite{anker}, whose return probabilities are given by (\ref{returnCH}) without the dominant exponential factor and give, therefore the spectral dimension function
\begin{equation} 
d_{\rm s} (t)  = D- (D-3) \left( {vJ t\over 1+vJ t} \right) \ .
\label{modspe}
\end{equation}
The spectral dimension on scales smaller than $1/J=R$ is $D$ and coincides with the topological dimension ($D=2$ in our case), while the large-scale spectral dimension at distances larger than $1/J=R$ becomes 3, independently of $D$. In the mathematical literature this is called the  ``pseudo-dimension", or ``dimension at infinity" of the constant negative curvature manifold \cite{anker}. The large-scale spectral dimension 3 is not confined to a constant negative curvature manifold though, but is valid for any manifold with strictly non-positive curvature and is a consequence of the central limit theorem in negative curvature \cite{cllt}. This scale-dependent spectral dimension reproduces similar results in causal dynamical triangulations \cite{cdt1, cdt2} and Horava-Lifshitz gravity \cite{horava}, with the difference that, here, there is a fundamental scale separating the two regimes. 

When we take into account the sub-dominant random component of negative-curvature Brownian motion two things happen: first, the universal time $t$ does not coincide anymore with one coordinate, secondly, an additional space dimension emerges spontaneously, a phenomenon we shall denote by hyperbolic holography. Let us consider the radial heat kernel $K(t, z)$ on negative curvature, for distances $Jz \gg1 $, such that the angular distribution is uniform, 
\begin{equation}
\left( \partial_\tau -\Delta \right) K(\tau, \rho) = 0 \ ,
\label{heat}
\end{equation}
where the radial Laplacian is given by
\begin{equation}
\Delta = \partial_\rho^2 + {\rm coth } \rho \ \partial_\rho \approx \Delta = \partial_\rho^2 + \partial_\rho \ ,
\label{radlap}
\end{equation}
and we have introduced, for simplicity of presentation, dimensionless quantities $\rho = Jz$ and $\tau = vJ t$. The kernel $K(\tau, \rho)$ admits the representation \cite{anker} 
\begin{eqnarray}
&&K(\tau, \rho) = {\rm e}^{-{1\over 4} \tau}\  f_0(\rho) \ k(\tau, \rho ) \ ,
\nonumber \\
&&\left( \Delta + {1\over 4} \right) f_0(\rho) = 0 \ ,
\label{sep}
\end{eqnarray}
where $f_0$ is the symmetric ground state of the Laplacian corresponding to the spectral gap \cite{davies}
\begin{equation}
f_0 (\rho) =(1+\rho) \ {\rm e}^{-{1\over 2} \rho} \ ,
\label{gs}
\end{equation}
for $\rho \gg1$. 
The residual kernel $k(\tau, \rho)$ describes the infinite Brownian loop random process, obtained by subtracting out the dominant ballistic component. It encodes the random dynamics of the defects as seen by an observer comoving with the inertial flow and satisfies the modified heat equation
\begin{equation} 
\left( \partial_\tau -\Delta - 2 \ \partial_\rho {\rm ln} f_0\left( \rho \right) \cdot \partial_\rho\right) k(\tau, \rho) = 0 \ ,
\label{ibl}
\end{equation}
Using (\ref{gs}) and $\rho \gg 1$, this equation becomes 
\begin{eqnarray}
\left( \partial_\tau -\left( \partial_\rho^2 +{2\over \rho} \partial_\rho \right) \right) k(\tau, \rho) &&= 0 \ ,
\nonumber \\
\left( \partial_\tau -\Delta_{3D} \right) k(\tau, \rho) &&= 0 \ ,
\end{eqnarray}
where $\Delta_{3D}$ is the radial Laplacian in a 3D Euclidean space. Therefore 
\begin{equation}
k(t, z) \asymp (vJ t)^{-{3\over 2}} {\rm e^{-{Jz^2 \over 4vt}}} \ ,
\label{iblkernel}
\end{equation}
where we have reintroduced the original variables. This is the isotropic heat kernel on a 3D Euclidean manifold with distance norm $z$ ``inherited" from the hyperbolic distance on the holographic screen \cite{anker}. In the Riemannian negative-curvature picture, the observer is comoving with the inertial flow of ballistic diffusion; in the Lorentzian positive-curvature picture, the observer is comoving with the Hubble flow on the holographic screen. If this Lorentzian picture is used for the dominant flow, the comoving random process must be treated as time-symmetric in the forward and backward directions, which gives immediately the Schr\"odinger evolution \cite{nelson}
\begin{equation}
K (t, \rho) \asymp \left( {m\over i\hbar  \tilde t} \right)^{3\over 2} \  {\rm e^{-{m z^2 \over 2i\hbar   \tilde t}}} \ ,
\label{qm}
\end{equation}
where we have used the previously introduced time $\tilde t$ measured in seconds. This is the 3D Schrödinger propagator for a particle of mass $m=\hbar J/ 2v$, velocity $v=J{\cal D}/2$ and therefore Compton wavelength $\lambda_{\rm C} =2 (v/c) (1/J) = 2(v/c) R$. This is the definition of the ratio $m/\hbar$, more about this in the next section. At large scales, the local random process, as seen by geodesically free-falling observers on a de Sitter holographic screen, is quantum mechanics in an effective 3D Euclidean space with distances given by the hyperbolic distance on the holographic screen. 

Quantum mechanics arises because the physical evolution parameter t, identified with coordinate time at large scales, must be treated as Lorentzian, when using the comoving interpretation of the dominant inertial flow. When constructing the Lorentz-invariant, but non-reparametrization-invariant action in terms of t, the fluctuations at small scales are sampled with the phase factor exp(iS). The Wick rotation is automatically enforced by the dominant comoving flow. The holographic promotion of the screen distance to an effective distance in a 3D Euclidean space immediately then implies quantum behaviour in this emergent space. The appearance of quantum behaviour, therefore goes hand-in-hand with the Lorentzian signature of space-time and vanishes with it below a scale fixed by the radius of curvature of the holographic screen.

\section{Emergent large-scale physics}

\subsection{Matter, the Einstein equations and the cosmological constant}
In the previous section we have derived the emergence of an effective 3D Euclidean space ``seen" at large distances by quantum particles in a comoving frame of a deSitter holographic screen. In this section we consider the holographic screen and the emergent space as embedded in a common large-scale bulk universe manifold. To this end let us consider the isometry groups of the two components. The isometry group of a constant-curvature 2D holographic screen is $SO(1,2)^{(+)}$, where the $+$ indicates the orthochronous subgroup if the curvature is negative, in which case an arrow of time is built in. The isometry group of the emergent 3D Euclidean space is $SO(3)$. The construction of an embedding universe must be such that these two isometry groups share a common $SO(2)$, corresponding to the angular variable of the holographic screen, and that 3D spatial distances scale with the same exponential factor as on the holographic screen, see (\ref{iblkernel}). These requirements are satisfied by an isometry group SO(1,4) \cite{gibbons, higuchi}, which corresponds, in the effective Lorentzian picture, to a (1+3)-dimensional de Sitter spacetime. The effective curvature experienced in this “bulk’’ de Sitter universe need not coincide with that of the holographic screen, since rescalings of time and of the speed of light can be absorbed into the definition of the de Sitter radius. Consequently, the non-relativistic quantum mechanics (\ref{qm}) can be ``lifted" to relativistic quantum field theory in an effective bulk de Sitter space. 

Let us now consider a random phase bubble embedded in a constant curvature graph at low, but finite coupling. The interior vertices of the bubble support no squares (apart accidental ones). The boundary vertices of the bubble also support typically less squares than the surrounding graph, since at least one edge has to connect them to the random graph inside. Moreover, the remaining long cycles based on edges of the boundary vertices are much longer than in the surrounding geometric phase, since they have to cycle along many vertices in the interior random graph before coming back. These energy and entropy effects conspire to generate the higher free energy of the random phase bubble with respect to the geometric surrounding. But also, each vertex on the boundary has a lower curvature since less squares and longer cycles imply a larger excess angle at the vertex, so that the curvature is more negative than in the geometric phase, as explained above. Since the number of boundary vertices depends on the bubble size, this lower negative curvature is associated with the higher free energy, but this relation is statistical in nature since many microstates of slightly different curvature, due to long cycles entropy,  characterize the ensemble with a given free energy. The ensemble equation is thus 
\begin{eqnarray}
-\left( \langle K\rangle- 4\Lambda_2 \right) \ell^2  = {\kappa \over \ell} F \ ,
\nonumber \\
\kappa =n(g)  {8\pi G_{\rm N} \over c^4} \ ,
\label{free1}
\end{eqnarray}
in units where the free energy $F$ is measured in Joule with respect to the surrounding geometric matrix with curvature $4\Lambda_2$. $G_{\rm N}$ is Newton's gravitational constant and $n(g)$ is a numerical factor depending on the original coupling $g$. 

In the Lorentzian picture the curvature changes sign. Let us consider also the large-scale continuum approximation in the bulk universe, where the random bubble becomes a point-like excitation at rest in the comoving frame. In this case we must consider density distributions by dividing both sides by $\ell^3$. This gives 
\begin{equation}
\left( \langle R(x) \rangle -4\Lambda_3 \right) = \kappa f(x) \ ,
\label{free2}
\end{equation}
where $f$ is the free energy density, measured here with respect to the bulk curvature $4\Lambda_3$, and we have used the standard notation $R(x)$ for the continuum Ricci curvature in (1+3) dimensions. Since both the free energy and the curvature of these random phase defects are higher than the surrounding manifold on which they live it is natural to identify them with matter particles, although they are made of the same microscopic degrees of freedom as the surrounding space. 

In a generic frame, the energy-momentum tensor of a dust of such particles moving along geodesics must be defined by a local (maximum two derivatives), symmetric geometric tensor which is covariantly conserved and has eq. (\ref{free2}) as its trace. The unique such geometric tensor is the Einstein tensor augmented by a cosmological constant term \cite{lovelock}, which gives
\begin{equation}
\langle R_{\mu \nu} \rangle -{1\over 2} \langle R \rangle g_{\mu \nu} + \Lambda_3 g_{\mu \nu} = \kappa T_{\mu \nu} \ ,
\label{einstein}
\end{equation}
These are Einstein's equations of general relativity (with a cosmological term), with the difference that $T_{\mu \nu}$ must be interpreted as a (density) thermodynamic potential and the equations hold as statistical ensemble averages. In this model these are not dynamical equations but, rather, macroscopic constitutive equations of matter as defined by the microscopic network degrees of freedom.

Matter, in this model is simply the random phase of the microscopic network. Matter particles have two sizes: a gravitational size $O \left( \ell_{\rm P}\right)$ where their associated excess curvature is concentrated and a quantum size, the Compton wavelength, set by the screen radius of curvature. In the present model, at large scales, it is ``geometry that defines matter", not the other way around, as usual.  

In this model, there is an obvious possible solution to the cosmological constant problem (for a review see, e.g. \cite{cosconst}): the zero-point energy of quantum fields curves only the holographic screen, only matter and geometric fluctuations curve the bulk universe. Note that the free energy of the model is, by construction, a very small positive quantity for a universe empty of matter, with only geometric fluctuations on the scale of the cutoff $\ell$.

\subsection{Black holes, Hawking radiation and absence of information paradox}
As discussed above, matter is modelled as the random phase of the network. At finite coupling $g < g_{\rm cr}$, matter particles are identified with the residual bubbles of random phase embedded in the dominant geometric matrix, representing the space manifold. Their typical size, the Planck scale, is stabilized by the balance between energy and entropy at the given coupling.  

These bubbles diffuse due to the underlying network fluctuations and, in the continuum approximation at large scales, their dominant dynamics in the Lorentzian picture is described by a thermodynamic energy momentum tensor defined by the Einstein equations statistically in terms of network curvature. As a consequence, matter particles will tend to aggregate under the influence of gravitational forces. This is balanced by repulsive effects, among which quantum ones, such as the uncertainty relation and statistical pressure for fermions, but also thermal and radiation pressure and other interactions. However, if many particles do manage to approach each other on distances of the Planck length order, their random networks can merge to form a larger bubble of random network. And if the resulting bubble becomes larger than the radius of curvature scale, we have a single macroscopic 3D quantum object. This is a black hole.

Actually the hole is not really black. Let us consider the typical times $\tau_{\rm G}$ and $\tau_{\rm R}$ necessary for information to traverse a geometric and a random region of the same size $L$. In the geometric phase $L \propto N_{\rm G}^{1/D}$, where $N_{\rm G}$ is the number of network vertices making up the region and $D=2$ on the screen or $D=3$ in the emergent space. In the random phase, instead, $L \propto {\rm ln}\  N_{\rm R}$. To achieve the same dimension $L$, we must choose $N_{\rm R} \propto {\rm exp} \ N_{\rm G}^{1/D}$. Therefore, to cover the same distance $L$ in the random phase, information must take exponentially more steps than in the geometric phase. If every edge is traversed with the same speed, this entails that the time needed to traverse a region of size $L$ of random phase is exponentially larger than the time it takes for the same distance on the geometric matrix,
\begin{equation}
\tau_{\rm R} \propto {\ell \over v} \ {\rm exp} \left({v \over \ell} \tau_{\rm G}\right) \ .
\label{char}
\end{equation}
The random phase acts as a retardant medium for information propagation due to the disorder associated with the lack of a coherent geometric matrix. From the point of view of an observer in the geometric matrix, clocks in the random phase tick exponentially slower. Albeit exponentially large, however, the time delay associated with traversing a large random phase bubble is still finite. After a very long time, information will come out of the random phase region; only in the continuum limit $N^{1/D} \to \infty$ and $\ell \to 0$ with $\ell N^{1/D}$ finite 
does the hole develop an event horizon and become really black, since information has to make infinite steps to come out. However, in this limit no random phase survives. True black holes are thus only a continuum idealization which does not exist at finite universe curvature. 

Note that k-regular ($k\ge 3$) random graphs are excellent expanders: the number of vertices within a ball of radius $r$ scales like the number of vertices at distance $r$, where $r$ denotes the level of hops along the tree leaves. Asymptotically for $r \gg 1$, the ``surface to volume ratio" of the graph is $k-2/k-1$, 2/3 for 4-regular graphs. This means that there is essentially no distinction between volume and surface since 2/3 of the volume is taken up by the boundary surface. Since the boundary is well-defined from the embedding in the geometric matrix, the black hole is essentially only a very thin surface layer. All the vertices of the random bubble are concentrated there. This is the origin of the holographic principle: the number of degrees of freedom of a black hole scales with the area of the (would be) event horizon. The holographic principle for black holes is thus an immediate consequence of the expander property of regular random graphs. 

The same expander property is shared by hyperbolic space. Since the geometric matrix of the screen is a negative-curvature surface, it is natural to visualize the black hole also as embedded in a 2D hyperbolic space, for which the surface to volume ratio is $1/R$, the inverse of the radius of curvature. To do so we must assign to the graph edges a fixed length unit $\ell_{\rm R}$ and then equate $1/R$ to $(2/3) 1/\ell_{\rm R}$, which gives a curvature radius $R=(3/2) \ell_{\rm R}$. Finally, if we want to compare this to the surrounding geometric matrix, we must rewrite the logarithmic distances in this hyperbolic bubble as $\ell_{\rm R} N_{\rm R}^{1/D}$ where $N_{\rm R}$ is the number of vertices in it and $\ell_{\rm R} =\ell \ {{\rm ln} \  N_{\rm R}\over N_{\rm R}^{1/D}}$. For large bubbles, the ``inside" effective edge length is exponentially smaller than the corresponding one in the geometric phase outside, which implies that the black hole can be seen as a hyperbolic bubble of exponentially large curvature with all degrees of freedom on the surface. 

The expander property of the random phase implies that black holes are essentially large circles of radius $R_{\rm BH}$ on the holographic screen. Their free energy, measured in Joule, can thus be expressed as
\begin{equation}
F = u {c^4\over 2G_N} R_{\rm BH} - g s {c^4\over 2G_N} R_{\rm BH}\ ,
\label{bh1}
\end{equation}
where $u$ and $s$ are numerical constants determining internal energy and entropy, respectively. We identify the internal energy as the black hole mass (times velocity squared),
\begin{equation}
M = u{c^2\over 2G_N} R_{\rm BH} \ ,
\label{bh2}
\end{equation}
which, for $u=1$ is the Schwarzschild value. Seen from afar, these black holes appear as 2D surfaces, since their scale is much larger than the screen radius of curvature. If we want to interpret correspondingly the entropy as distributed on this surface, we must multiply it by a factor $4\pi R_{\rm BH}$ and divide simultaneously the dimensionless ``temperature" g by the same factor. This gives the physical temperature and entropy seen in the bulk, 
\begin{eqnarray} 
T_{\rm BH} &&= g {c^3 \hbar \over 8\pi k_{\rm B} G_{\rm N} M_{\rm BH}} \ ,
\nonumber \\
S_{\rm BH} &&= 2su {c^3 k_{\rm B}\over 4 \hbar G_{\rm N}} A_{\rm BH} \ ,
\label{bh3}
\end{eqnarray}
where $A_{\rm BH}$ is the area of the 2D surface.
For $g=1$ and $su=1/2$ these are the Hawking temperature \cite{hawking} and the Bekenstein entropy \cite{bekenstein}. The exact values of these numerical factors depend, of course, on the microscopic statistical mechanics of the underlying network. However, the qualitative facts that black holes are characterized by an entropy proportional to their perceived horizon area and by a temperature inversely proportional to their mass are natural, but non-trivial predictions of the network model. 

Bubbles of random phase much larger than the Planck length are not stable since they are not minima of the network free energy. When such a large bubble forms it will stabilize by shrinking to its equilibrium size by emitting both energy and entropy. This is a simple explanation of the Hawking radiation of black holes. They are simply statistically unstable and they emit energy and entropy to return to their equilibrium size. 

Finally, since information is always stored in the fundamental network, there is no information paradox. When information falls into the black hole it is simply stored provisionally in another network phase and then released back when the hole stabilizes to its equilibrium size.

\subsection{Dark matter}
Dark matter is one of the major puzzles of contemporary cosmology (for a review see, e.g. \cite{arbey}). The mass-energy of the universe is made only by 5\% of ordinary baryonic matter, the remainder being 27\% dark matter, an unknown form of matter interacting with ordinary matter only through gravity and 68\% dark energy, attributed typically to the cosmological constant discussed above. 

There are two possible avenues to tackle the dark matter problem. One is to posit a new type of particle, with essentially only gravitational interactions, which is the mainstream belief today, the other is to posit a modification of classical general relativity (GR). In \cite{verlinde2, dark}, however, it was proposed that dark matter is a macroscopic manifestation of quantum gravity. We now focus on how dark matter naturally fits in the present model. 

We have shown that the holographic graph can have two phases, a random phase, identified with ordinary matter,  in which there are no short loops and a geometric phase, representing space, in which each vertex $i$ can exist in four possible states, corresponding to the number $\square_i=1,2,3,4$ of squares surrounding it. The number $\square_i$ determines the quantized curvature at this vertex as shown above. Until now we have focused on uniform curvature configurations in which all vertices in the geometric phase have the same number $\square$ of squares and we have shown that this curvature is inherited in the emergent (1+3)-dimensional space-time. 

What about inhomogeneous space configurations with different values of $\square_i$? Imagine, in particular that the coupling $g$ decreases from the critical coupling $g_{\rm cr}$ so that the universe emerges from an all-random state at an extremely high curvature $1/\ell^2$, representing the big bang. As always when lowering the coupling in continuous phase transitions, there typically survive domains of higher free energy. These are metastable states, which, depending on their free energy difference to the minimum and the barrier in between, may be extremely long-lived. These states are called allotropes and the most famous example are the two states of carbon, graphite, the free energy minimum and diamond the higher free energy allotrope (for a review see \cite{bernstein}). 

\begin{figure}[t!]
\includegraphics[width=7cm]{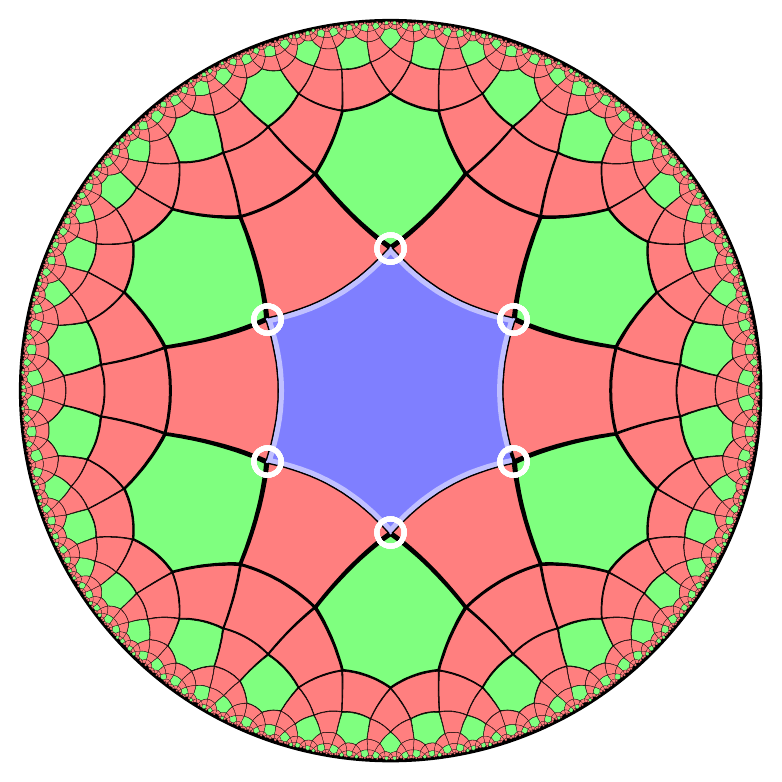}
\vspace{-0.3cm}
\caption{An allotropy region (blue) with vertices surrounded by two squares within an hyperbolic tessellation with all vertices surrounded by three squares. The hyperbolic tessellation represents space of a given uniform curvature; the blue allotropy region represents a region of dark matter, with a higher absolute curvature and free energy. For simplicity of presentation we are showing here only the smallest example of such a region. Created by Eryk Kopczyński using RogueViz \cite{marek}.}
\label{fig:Fig.9}
\end{figure}

In the present case, we can have metastable domains characterized by a higher number $\square_{\rm dom}$ than the prevailing configuration $\square$ and therefore having
a larger free energy than the space around them at a given $g$. Because of this higher free energy they have also a higher curvature (in the Lorentzian picture) and interact gravitationally. But they are neither empty space nor ordinary matter. These domains are natural candidates to represent dark matter in the universe. Dark matter, in this model, appears like crystal allotropy in the network fabric of space-time. An example is shown in Figure \ref{fig:Fig.9}.

Of course, for couplings $g$ midway between values favouring two different number of squares as the free energy minima, small domains of lower absolute curvature can form before the whole space-time transitions to this configuration. These would correspond to antigravity domains within a uniform metastable universe.


\begin{thebibliography}{10}
	
\bibitem{kiefer} C. Kiefer, Quantum Gravity, Oxford University Press, Oxford (2007). 

\bibitem{kuchar} K. V. Kuchar, Time and interpretations of quantum gravity, {\it Int. J. Mod. Phys.} {\bf D20} 3 (2011). 

\bibitem{horava} P. Horava, Quantum gravity at a Lifshitz point, {\it Phys. Rev. } {\bf D79} 084008 (2009). 

\bibitem{polchinski} J. Polchinski, String Theory, Cambridge University Press, Cambridge (UK (1998). 

\bibitem{as1} A. Eichhorn, Asymptotically safe gravity, proceedings of the 57th Course of the Erice International School of Subnuclear Physics, "In search for the unexpected", June 2019, arXiv: 2003.00044, (2019). 

\bibitem{as2} A. Eichhorn, The microscopic structure of quantum space-time and matter from a renormalization group perspective, {\it Nat. Phys.} {\bf 19} 1527-1529 (2023). 

\bibitem{loopqg} A. Ashtekar and E. Bianchi, A short review of loop quantum gravity, {\it Rep. Prog. Phys} {\bf 84} 042001 (2021). 

\bibitem{cdt1} 
J. Ambjorn, J. G\"orlich, J. Jurkiewicz and R. Loll, Nonperturbative Quantum Gravity, {\it Phys. Rep. } {\bf 519} 127-210 (2012) 

\bibitem{cdt2} R. Loll, Quantum Gravity from Causal Dynamical Triangulations: A Review, {\it Classical and Quantum Gravity}  {\bf 37} 013002 (2019). 

\bibitem{oriti} D. Oriti, The microscopic dynamics of quantum space as a group field theory. In: Foundations of space and time: reflections on quantum gravity. Aug 10–14 2009; Cape Town, South Africa; Cambridge University Press  257 (2011). 

\bibitem{tensor1} J. Ambjorn, B. Durhuus and T. Jonsson, Three-dimensional simplicial quantum gravity and generalized matrix models, {\it Mod. Phys. Lett.} {\bf A06} 1133-1146 (1991).

\bibitem{tensor2} N. Sasakura, Tensor model for gravity and orientability of manifold, {\it Mod. Phys. Lett.} {\bf A06} 2613-2624 (1991). 

\bibitem{sasakura} N. Sasakura, Phase profile of the wave function of canonical tensor model and emergence of large spacetimes, {\it Int. Jour. Mod. Phys.} {\bf 36} 2150222 (2021). 

\bibitem{causalsets} S. Surya, The causal set approach to quantum gravity, {\it Living Reviews in Relativity} {\bf 22:5} (2019). 

\bibitem{gorard} J. Gorard, Some Relativistic and Gravitational Properties of the Wolfram Model, {\it Complex Systems} {\bf 29} 599-654 (2020). 

\bibitem{verlinde1} E. Verlinde, On the origin of gravity and the laws of Newton, {\it JHEP} {\bf 1104} 029 (2011). 

\bibitem{verlinde2} E. Verlinde, Emergent gravity and the dark universe, {\it SciPost Phys.} {\bf 2} 016 (2017).

\bibitem{bianconient} G. Bianconi, Gravity from entropy, {\it Phys. Rev. D} {\bf 111} 066001 (2025).

\bibitem{wheeler} J. A. Wheeler, Information, physics, quantum: the search for links, Proceedings of the III international symposium on the foundations of quantum mechanics, 354-358, Tokyo (1989) . 

\bibitem{zanardi}S. Garneroni, P. Giorda and P. Zanardi, Bipartite quantum states and random complex networks, {\it New Journal of Physics} {\bf 14} (2012) 013011. 


\bibitem{forman1} R. Forman, Combinatorial Novikov-Morse theory, {\bf Int. J. Math.} {\bf 13} 333 (2002).

\bibitem{forman2} R. Forman, Bochner's Method for Cell Complexes and Combinatorial Ricci Curvature, {\it Discrete Comput. Geom.} {\bf 29} 323 (2003). 

\bibitem{olli0} Y. Ollivier, Ricci curvature of metric spaces, {\it C. R. Math. Acad. Sci. Paris} {\bf 345} 643-646 (2007). 

\bibitem{olli1}Y. Ollivier, Ricci curvature of Markov chains in metric spaces, {\it J. Funct. Anal.} {\bf 256} (2009) 810; 

\bibitem{olli2} Y. Ollivier, A survey of Ricci curvature fo metric spaces and Markov chains, {\it Adv. Stud. Pure Math.} {\bf 57} (2010) 343-381 (2010). 

\bibitem{olli3}Y. Lin, L. Lu and S. T. Yau, Ricci curvature of graphs, {\it Tohoku Math. J.} {\bf 63} (2011) 605-627. 

\bibitem{olli4} J. Jost and S. Liu, Ollivier's Ricci curvature, local clustering and curvature dimension inequalities on graphs, {\it Discrete Comput. Geom.} {\bf 51} (2014) 300-322. 

\bibitem{comb1} C. A. Trugenberger, Combinatorial quantum gravity: geometry from random bits,  {\it JHEP} {\bf 09} 045 (2017).

\bibitem{comb2} C. Kelly, C. A. Trugenberger, and F. Biancalana, Self-Assembly of Geometric Space from Random Graphs, {\it Classical and Quantum Gravity} {\bf 36} 125012 (2019). 

\bibitem{comb3} C. Kelly, C. A. Trugenberger and F. Biancalana, Emergence of the circle in a statistical model of random cubic graphs, {\it Classical and Quantum Gravity} {\bf 38} 075008 (2021). 

\bibitem{graphrev}R. Albert and L. Barabasi, Statistical mechanics of complex networks, {\it Rev. Mod. Phys.} {\bf 74} (2002) 47. 


\bibitem{bianconi2} G. Bianconi and C. Rahmede, Network geometry with flavor: from complexity to quantum geometry, {\it Phys. Rev.} {\bf E93} 032315 (2016). 

\bibitem{bianconi3} G. Bianconi and C. Rahmede, Emergent hyperbolic network geometry, {\it Sci. Rep.} {\ bf 7} 41974 (2017). 

\bibitem{universe} C. A. Trugenberger, Combinatorial quantum gravity and emergent 3D quantum behaviour, {\it Universe} {\bf 9} 499 (2023). 

\bibitem{dark} C. A. Trugenberger, Dark matter and dark energy in combinatorial quantum gravity, {\it Classical and Quantum Gravity} {\bf 41} 217002 (2024). 

\bibitem{spectralcdt} J. Ambjorn, J. Jurkiewicz and R. Loll, Spectral dimension of the universe, {\it Phys. Rev. Lett.} {\bf 95} 171301 (2005). 

\bibitem{reuter} O. Lauscher and M. Reuter, Fractal spacetime structure in asymptotically safe gravity, {\it JHEP} {\bf 05} 10:050 (2005). 

\bibitem{thooft}G. 't Hooft, Dimensional reduction in quantum gravity, arXiv:gr-qc/9310026.

\bibitem{susskind} L. Susskind, The world as a hologram, {\it J. Math. Phys.} {\bf 36} 6377-6396 (1995). 

\bibitem{bousso} R. Bousso, The holograpic principle, {\it Rev. Mod. Phys.} {\bf 74} 825-874 (2002). 

\bibitem{binder}K. Binder, D. Stauffer and H. M\"uller-Krumbhaar, Theory for the dynamics of clusters near the critical point. I. Relaxatlon of the Glauber kinetic Ising model, {\it Phys. Rev.} {\bf B12} 5361 (1975). 

\bibitem{tersenghi} C. Godr\`eche, F. Krzakala and F. Ricci-Tersenghi, Nonequilibrium critical dynamics of the ferromagnetic Ising model with Kawasaki dynamics, {\it J. Stat. Mech.: Theor. Exp. } {\bf P04007} (2004). 

\bibitem{luck} C. Godr\`eche and J. M. Luck, Anomalous self-diffusion in the ferromagnetic Ising chain with Kawasaki dynamics, {\it J. Phys. } {\bf A36} 9973-9980 (2003). 

\bibitem{prat} J.-J. Prat. Etude asymptotique et convergence angulaire du mouvement brownien sur une vari\'et\'e \`a courbure n\'egative. {\it C. R. Acad. Sci. Paris S\'er. A-B} {\bf 280(22):Aiii} A1539-A1542 (1975). 

\bibitem{kendall} W. S. Kendall, Brownian motion on 2-dimensional manifolds of negative curvature. {\it Trans. Amer. Math. Soc.} {\bf 275} 751-760 (1983). 

\bibitem{hsu} P. Hsu and W. S. Kendall, Limiting angle of Brownian motion in certain two-dimensional Cartan-Hadamard manifolds {\it Annales de la facult\'e des Sciences de Toulouse} {\bf 1} 169-186 (1982). 

\bibitem{hsubrief} E. P. Hsu, A brief introduction to Brownian motion on a Riemann manifold, Summer School in Kyushu (2008). 

\bibitem{hsubook} E. P. Hsu, Stochastic analysis on manifolds, {\it Graduate studies in mathematics} {\bf 38}, Providence (RI) (2002). 

\bibitem{arnaudon}M.Arnaudon and A. Thalmeier, Brownian motion and negative curvature, {\it Progress in Probability} {\bf 64} 145-163 (2011). 

\bibitem{eff1} C. A. Trugenberger, Emergent time, cosmological constant and boundary dimension at infinity in combinatorial quantum gravity, {\it JHEP} {\bf 04} 019 (2022). 

\bibitem{eff2} C. A. Trugenberger , Effective de Sitter space, quantum behaviour and large-scale spectral dimension (3+1), {\it JHEP} {\bf 03} 186 (2023). 

\bibitem{anker} J.-P. Anker, P. Bougerol and T. Jeulin, The infinite Brownian loop on a symmetric space, {\it Rev. Mat. Iberoamericana} {\bf 18} 41-97 (2002). 

\bibitem{ledrappier}F. Ledrappier, Central limit theorem in negative curvature {\it Ann. Probab.} {\bf 23} 12119-1233 (1995). 

\bibitem{cllt} F. Ledrappier and S. Lim, Local limit theorem in negative curvature, {\it Duke Math. J.} {\bf 170} 1585-1681 (2021). 

\bibitem{davies} E. B. Davies and N. Mandouvalos, Heat Kernel bounds on hyperbolic space and Kleinian groups, {\it Proc. London Math. Soc. (3) } {\bf 52} 182-208 (1988).

\bibitem{polyakovhouches} A. Polyakov, Two-dimensional quantum gravity, in ``Fields, Strings and Critical Phenomena", E. Br\`ezin and J. Zinn-Justin eds., Les Houches 1988, North Holland, Amsterdam (NL) (1990). 

\bibitem{newman}J. Park and M. E. J. Newman, Statistical mechanics of networks, {\it Phys. Rev. } {\bf E70} 066117 (2004). 

\bibitem{metric}D. P. Dailey, On the graphical containment of discrete metric spaces, {\it Discrete Mathematics} {\bf 131} 51-66 (1994). 

\bibitem{regge} T. Regge, General relativity without coordinates, {\it Nuovo Cimento} {\bf 19} 558-571 (1961). 

\bibitem{bochner} S. Bochner, Vector fields and Ricci curvature, {\it Bull. Amer. Math. Soc.} {\bf 52} 737-775 (1946). 

\bibitem{hatcher} A. Hatcher, Algebraic Topology, Cambridge University Press(2002).

\bibitem{sreejith} R. P. Sreejith, K. Mohanraj, J. Jost, E. Saucan and A. Samal, Forman curvature for complex networks, {\it J. Stat. Mech.: Theory Exp.} {\bf 2016}  063206 (2016).

\bibitem{samal} A. Samal, R. Sreejith, J. Gu, S. Liu, E. Saucan and J. Jost, Comparative analysis of two discretizations of Ricci curvature for complex networks, {\it Sci. Rep.}, {\bf 8 }1 (2018).

\bibitem{tee} P. Tee and C. A. Trugenberger, Enhanced Forman curvature and its relation to Ollivier curvature, {\it EPL} {\bf 133} 60006 (2021). 

\bibitem{klit1} N. Klitgaard and R. Loll, Introducing Quantum Ricci Curvature, {\it Phys. Rev.}  {\bf D97} 046008 (2018). 

\bibitem{klit2} N. Klitgaard and R. Loll, Implementing Quantum Ricci Curvature, {\it Phys. Rev.} {\bf D97} 106017 (2018). 

\bibitem{klit3} N. Klitgaard and R. Loll, How round is the quantum de Sitter universe? {\it Eur. Phys. J.} {\bf 80} 990 (2020). 

\bibitem{vanHoorn}P. van der Hoorn, W. Cunningham, G. Lippner, C. A. Trugenberger and D. Krioukov, Ollivier-Ricci curvature convergence in random geometric graphs, {\it Phys. Rev. Res.} {\bf 3} 013211 (2021). 

\bibitem{gilbert}E. N. Gilbert, Random Plane Networks, {\it Journal of the Society for Industrial and Applied Mathematics} {\bf 9} 533–543 (1961).

\bibitem{penrose} R. Penrose, Random geometric graphs, Oxford Universit Press, Oxford (UK) (2003). 

\bibitem{kahle}M. Kahle, Random geometric complexes, {\it Discrete Comput. Geom.} {\bf 45} 553 (2011).

\bibitem{latsche} J. Latschev, Vietoris-Rips complexes of metric spaces near a closed Riemannian manifold, {\it Arch. Math.} {\bf  77} 522 (2001).

\bibitem{bakry} D. Cushing, S. Liu and N. Peyerimhoff, Bakry-Emery curvature functions on graphs, {\it Canad. J. Math.} {\bf 72} 89-143 (2020). 

\bibitem{resistance} K. Devriendt and R. Lambiotte, Discrete curvature on graphs from the effective resistance, {J. Phys. Complex.} {\bf 3} 025008 (2022). 

\bibitem{cdthl}J. Ambjørn, L. Glaser, Y. Sato and Y. Watabiki, 2d CDT is 2d Horava-Lifshitz quantum gravity,  {\it Phy. Lett.} {\bf B722} 172-175 (2013). 

\bibitem{evnin} P. Akara-Pipattana, T. Chotibut and O. Evnin, The birth of geometry in exponential random graphs, {\it J. Phys. } {\bf A54} 425001 (2021). 


\bibitem{gorsky1}A. Gorsky and O. Valba, Interacting thermofield doubles and critical behaviour in random regular graphs, {\it Phys. Rev.} {\bf D103} 106013 (2021). 

\bibitem{gorsky2} A. Gorsky, V. Kazakov, F. Levkovich-Maslyuk and V. Mishnyakov, A flow in the forest, {\it JHEP} {\bf 03} 067 (2023). 

\bibitem{krioukov} D. Krioukov, Clustering implies geometry in networks, {\it Phys. Rev. Lett.} {\bf 19} 208302 (2016). 

\bibitem{oneil}P. E. O’Neil, Asymptotics in random (0,1)-matrices, {\it Bull. Am. Math. Soc.} {\bf 75} 1276 (1969). 

\bibitem{genus} C. Thomassen, Embeddings of graphs, {\it Discrete Mathematics} {\bf 124} 217-228 (1994). 

\bibitem{expander}H. Namazi, P. Pankka and J. Souto, Distributional limits of Riemannian manifolds and graphs with sublinear genus growth, {\it Geometric and Functional Analysis} {\bf 24} 322-359 (2014). 

\bibitem{seiberg} N. Seiberg, Notes on quantum Liouville theory and quantum gravity, {\it Rev. Mod. Phys.} {\bf 102} 319-349 (1990). 

\bibitem{tasi}P. Ginsparg and G. Moore, Lectures on 2D Gravity and 2D String Theory, TASI Summer School 1992, hep-th/9304011. 

\bibitem{shiga} K. Shiga, Hadamard manifolds, in ``Geometry of geodesics and related topics", {\it Advanced Studies in Pure Mathematics} {\bf 3} 239-281 (1984). 

\bibitem{anderson} J. W. Anderson, Hyperbolic geometry, Springer-Verlag, Berlin (2005). 

\bibitem{datta} B. Datta and S. Gupta, Semi-regular tilings of the hyperbolic plane, {\it Discrete Comput. Geom.}  {\bf 65} 531 (2019).

\bibitem{maiti1} A. Maiti, Quasi-vertex-transitive maps on the plane, {\it Discrete Math.} {\bf 343}, 7 (2020). 
	
\bibitem{maiti2} A. Maiti, Pseud-homogeneous tilings of the hyperbolic plane, arxiv:2302.05661 (2023).

\bibitem{strominger} M. Spradlin, A. Strominger and A. Volovich, Les Houches lectures on de Sitter space, arXiv:hep-th/0110007. 	

\bibitem{grigorian} A. Grigor\' yan, Estimates of heat kernels on Riemannian manifolds, in ``Spectral Theory and Geometry", Cambridge University Press, Cambridge (2010). 

\bibitem{nelson} E. Nelson, Derivation of the Schr\"odinger equation form Newtonian mechanics, {\it Phys. Rev.} {\bf 150} 1079 (1966). 

\bibitem{gibbons} G. W. Gibbons and S. W. Hawking, Cosmological event horizon thermodynamics and particle creation, {\it Phys. Rev.} {\bf D15} 2738 (1977).

\bibitem{higuchi} A. Higuchi, Symmetric tensor spherical harmonics on N-sphere and their application to the de Sitter group, {\it J. Math. Phys.} {\bf 28} 1553 (1987). 

\bibitem{lovelock} D. Lovelock, The Einstein tensor and its generalizations, {\it J. Math. Phys.} {\bf 12} 498-501 (1971). 

\bibitem{cosconst} T. Padmanabhan, Cosmological constant-the weight of the vacuum, {\it Phys. Rept.} {\bf 380} 235-320 (2003). 

\bibitem{hawking} S. W. Hawking, Particle creation by black holes, {\it Comms. Math. Phys.} {\bf 43} 199 (1975).

\bibitem{bekenstein} J. D. Bekenstein, Black holes and entropy, {\it Phys. Rev. } {\bf D7} 2333 (1973). 

 \bibitem{arbey}A. Arbey, Dark matter and the early universe: a review, {\it Progress in Particle and Nuclear Physics} {\bf 119} 103865, (2021). 

\bibitem{bernstein}J. Bernstein, Polymorphism in molecular crystals, Oxford University Press, Oxford (2002). 

 \bibitem{marek}E. Kopczynski, D. Celinska and M. Ctranct, HyperRogue: playing with hyperbolic geometry, Bridges Conference Proceedins (2017) 
 
	

	

        










 




























	
	
\end{thebibliography}
\end{document}